\documentclass[preprint]{aastex62}
\usepackage{subfigure}
\usepackage{amsmath}
\usepackage{xcolor}
\usepackage{amssymb}

\newcommand{\feh}{\hbox{$ [\mathrm{Fe}/\mathrm{H}]$}} 
 
\newcommand{\alphafe}{\hbox{$ [\mathrm{\alpha}/\mathrm{Fe}]$}}

\newcommand{\Mh}{\hbox{$ [{\rm M}/{\rm H}]$}}

\newcommand{\kms}{\hbox{$\mathrm{km\,s}^{-1}$}}
\newcommand{\SNR}{\hbox{SNR}}
\newcommand{\dex}{\hbox{dex}}

\newcommand{\logg}{\hbox{$\log g$}}
\newcommand{\teff}{\hbox{$T_\mathrm{eff}$}}
\newcommand{\rave}{\textsc{Rave}}
\newcommand{\ie}{i.e.}
\newcommand{\eg}{e.g.}

\shorttitle{\rave\ DR6 - I.: spectra and radial velocities}
\shortauthors{Steinmetz et al.}

\begin{document}
\title{The Sixth Data Release of the Radial Velocity Experiment (\rave) -- I: Survey Description, Spectra and Radial Velocities}

\author[0000-0001-6516-7459]{Matthias Steinmetz}
\affiliation{Leibniz-Institut f{\"u}r Astrophysik Potsdam (AIP), An der Sternwarte 16, 
14482 Potsdam, Germany}

\author[0000-0002-6070-2288]{Gal Matijevi\v c}
\affiliation{Leibniz-Institut f{\"u}r Astrophysik Potsdam (AIP), An der Sternwarte 16, 14482 Potsdam, Germany}

\author[0000-0002-2366-8316]{Harry Enke}
\affiliation{Leibniz-Institut f{\"u}r Astrophysik Potsdam (AIP), An der Sternwarte 16, 14482 Potsdam, Germany}

\author[0000-0002-2325-8763]{Toma\v{z} Zwitter}
\affiliation{University of Ljubljana, Faculty of Mathematics and Physics, Jadranska 19, 
SI-1000 Ljubljana, Slovenia}

\author[0000-0002-1317-2798]{Guillaume Guiglion}
\affiliation{Leibniz-Institut f{\"u}r Astrophysik Potsdam (AIP), An der Sternwarte 16, 
14482 Potsdam, Germany}

\author[0000-0002-8861-2620]{Paul J. McMillan}
\affiliation{Lund Observatory, Department of Astronomy and Theoretical Physics, Lund 
University, Box 43, 22100 Lund, Sweden}

\author[0000-0002-9035-3920]{Georges Kordopatis}
\affiliation{Universit{\'e} C{\^o}te d'Azur, Observatoire de la C{\^o}te d'Azur, CNRS, 
Laboratoire Lagrange, France}

\author[0000-0003-0974-4148]{Marica Valentini}
\affiliation{Leibniz-Institut f{\"u}r Astrophysik Potsdam (AIP), An der Sternwarte 16, 
14482 Potsdam, Germany}

\author[0000-0003-1269-7282]{Cristina Chiappini}
\affiliation{Leibniz-Institut f{\"u}r Astrophysik Potsdam (AIP), An der Sternwarte 16, 
14482 Potsdam, Germany}

\author[0000-0003-2688-7511]{Luca Casagrande}
\affiliation{Research School of Astronomy \& Astrophysics, The Australian National University, Canberra, Australia}

\author[0000-0002-3233-3032]{Jennifer Wojno}
\affiliation{The Johns Hopkins University, Department of Physics and Astronomy, 3400 N. 
Charles Street, Baltimore, MD 21218, USA}

\author[0000-0001-5261-4336]{Borja Anguiano} 
\affiliation{Department of Astronomy, University of Virginia, Charlottesville, VA, 22904, USA}

\author[0000-0002-4605-865X]{Olivier Bienaym\'e} \affiliation{Observatoire astronomique de Strasbourg, Universit\'e de Strasbourg, CNRS, 
11 rue de l'Universit\'e, F-67000 Strasbourg, France }

\author{Albert Bijaoui}
\affiliation{Universit{\'e} C{\^o}te d'Azur, Observatoire de la C{\^o}te d'Azur, CNRS, 
Laboratoire Lagrange, France}

\author{James Binney}
\affiliation{Rudolf Peierls Centre for Theoretical Physics, Clarendon Laboratory, Parks Road, Oxford, OX1 3PU, UK}

\author[0000-0001-6516-7459]{Donna Burton}
\affiliation{Australian Astronomical Observatory, Siding Spring, Coonabarabran NSW 2357, Australia}
\affiliation{University of Southern Queensland (USQ), West Street Toowoomba Qld 4350 Australia}

\author{Paul Cass}
\affiliation{Australian Astronomical Observatory, Siding Spring, Coonabarabran NSW 2357, Australia}

\author{Patrick de Laverny}
\affiliation{Universit{\'e} C{\^o}te d'Azur, Observatoire de la C{\^o}te d'Azur, CNRS, 
Laboratoire Lagrange, France}

\author{Kristin Fiegert}
\affiliation{Australian Astronomical Observatory, Siding Spring, Coonabarabran NSW 2357, Australia}

\author[0000-0001-6280-1207]{Kenneth Freeman}
\affiliation{Research School of Astronomy \& Astrophysics, The Australian National University, Canberra, Australia}

\author{Jon P. Fulbright}
\affiliation{The Johns Hopkins University, Department of Physics and Astronomy, 3400 N. Charles Street, Baltimore, 
MD 21218, USA}

\author[0000-0003-4446-3130]{Brad K. Gibson}
\affiliation{E.A. Milne Centre for Astrophysics, University of Hull, Hull, HU6 7RX, 
United Kingdom}

\author[0000-0003-4632-0213]{Gerard Gilmore}  
\affiliation{Institute of Astronomy, Cambridge, UK}

\author[0000-0002-1891-3794]{Eva K.\ Grebel}
\affiliation{Astronomisches Rechen-Institut, Zentrum f\"ur Astronomie
der Universit\"at Heidelberg, M\"onchhofstr.\ 12--14, 69120 Heidelberg,
Germany}

\author[0000-0003-3937-7641]{Amina Helmi}  
\affiliation{Kapteyn, Astronomical Institute, University of Groningen, P.O. Box 800, 9700 AV Groningen, The Netherlands}

\author[0000-0002-2808-1370]{Andrea Kunder}  
\affiliation{Saint Martin's University, 5000 Abbey Way SE, Lacey, WA, 98503, USA}

\author[0000-0001-6805-9664]{Ulisse Munari}
\affiliation{INAF Astronomical Observatory of Padova, 36012 Asiago (VI), Italy}

\author{Julio F. Navarro}
\affiliation{Department of Physics and Astronomy, University of Victoria, Victoria, BC, 
Canada V8P5C2.}

\author[0000-0001-6516-7459]{Quentin Parker}
\affiliation{CYM Physics Building, The University of Hong Kong, Pokfulam, Hong Kong SAR, PRC}
\affiliation{The Laboratory for Space Research, Hong Kong University, 
Cyberport 4, Hong Kong SAR, PRC}

\author{Gregory R. Ruchti}
\altaffiliation{deceased}
\affiliation{The Johns Hopkins University, Department of Physics and Astronomy, 3400 N. Charles Street, Baltimore, 
MD 21218, USA}

\author{Alejandra Recio-Blanco}
\affiliation{Universit{\'e} C{\^o}te d'Azur, Observatoire de la C{\^o}te d'Azur, CNRS, 
Laboratoire Lagrange, France}

\author{Warren Reid}
\affiliation{Department of Physics and Astronomy, Macquarie University, Sydney, NSW 
2109, Australia}
\affiliation{Western Sydney University, Locked bag 1797, Penrith South, NSW 2751, 
Australia}

\author[0000-0003-4072-9536]{G. M. Seabroke}
\affiliation{Mullard Space Science Laboratory, University College London, Holmbury St Mary, Dorking, RH5 6NT, UK}

\author{Alessandro Siviero}
\affiliation{Dipartimento di Fisica e Astronomia G. Galilei, Universita' di Padova, 
Vicolo dell'Osservatorio 3, I-35122, Padova, Italy}

\author{Arnaud Siebert}
\affiliation{Observatoire astronomique de Strasbourg, Universit\'e de Strasbourg, CNRS, 
11 rue de l'Universit\'e, F-67000 Strasbourg, France }

\author{Milorad Stupar}  
\affiliation{Australian Astronomical Observatory, Siding Spring, Coonabarabran NSW 2357, Australia}
\affiliation{Western Sydney University, Locked Bag 1797, Penrith South, NSW 2751, Australia}

\author[0000-0002-3590-3547]{Fred Watson}  
\affiliation{Department of Industry, Innovation and Science, 105 Delhi Rd, North Ryde, NSW 2113, Australia}

\author{Mary E.K.\ Williams}
\affiliation{Leibniz-Institut f{\"u}r Astrophysik Potsdam (AIP), An der Sternwarte 16, 
14482 Potsdam, Germany}

\author[0000-0002-4013-1799]{Rosemary F.G.\ Wyse}
\affiliation{The Johns Hopkins University, Department of Physics and Astronomy, 3400 N. Charles Street, Baltimore, 
MD 21218, USA}
\affiliation{Kavli Institute for Theoretical Physics, University of California, Santa Barbara, CA 93106, USA}

\author[0000-0003-4524-9363]{Friedrich Anders}
\affiliation{Institut de Ci\`encies del Cosmos, Universitat de Barcelona (IEEC-UB), 
Mart\'i i Franqu\`es 1, 08028 Barcelona, Spain}
\affiliation{Leibniz-Institut f{\"u}r Astrophysik Potsdam (AIP), An der Sternwarte 16, 
14482 Potsdam, Germany}

\author[0000-0003-2595-5148]{Teresa Antoja} 
\affiliation{Institut de Ci\`{e}ncies del Cosmos de la Universitat de Barcelona, Mart\'{i} i Franqu\`{e}s 1, 09028 Barcelona (Spain)}

\author[0000-0002-5571-5981]{Danijela Birko}
\affil{University of Ljubljana, Faculty of Mathematics and Physics, Ljubljana, Slovenia}

\author[0000-0001-7516-4016]{Joss Bland-Hawthorn}
\affiliation{Sydney Institute for Astronomy, School of Physics,
The University of Sydney, NSW 2006, Australia}

\author[0000-0002-9480-8400]{Diego Bossini}
\affiliation{Instituto de Astrof\'isica e Ci$\hat{e}$ncias do Espa\c{c}o, Universidade do Porto, CAUP, Rua das Estrelas, 4150-762 Porto,Portugal} 

\author[0000-0002-8854-3776]{Rafael A. Garc\'\i a} 
\affiliation{IRFU, CEA, Universit\'e Paris-Saclay, F-91191 Gif-sur-Yvette, France} 
\affiliation{AIM, CEA, CNRS, Universit\'e Paris-Saclay, Universit\'e Paris Diderot, Sorbonne Paris Cit\'e, F-91191 Gif-sur-Yvette, France} 

\author[0000-0002-0759-0766]{Ismael Carrillo} 
\affiliation{Leibniz-Institut f{\"u}r Astrophysik Potsdam (AIP), An der Sternwarte 16, 
14482 Potsdam, Germany}

\author[0000-0002-5714-8618]{William J. Chaplin} 
\affiliation{School of Physics and Astronomy, University of Birmingham, Edgbaston, Birmingham B15 2TT, UK}
\affiliation{Stellar Astrophysics Centre (SAC), Department of Physics and Astronomy, Aarhus University,  DK-8000 Aarhus C, Denmark} 

\author{Yvonne Elsworth}
\affiliation{School of Physics and Astronomy, University of Birmingham, Edgbaston, Birmingham, B15 2TT, UK}
\affiliation{Stellar Astrophysics Centre (SAC), Department of Physics and Astronomy, Aarhus University,  DK-8000 Aarhus C, Denmark}

\author[0000-0003-3180-9825]{Benoit Famaey}
\affiliation{Observatoire astronomique de Strasbourg, Universit\'e de Strasbourg, CNRS, 11 rue de l'Universit\'e, F-67000 Strasbourg, France }

\author[0000-0003-3333-0033]{Ortwin Gerhard}
\affiliation{Max-Planck-Institut f{\"u}r extraterrestrische Physik, Postfach 1312,
  Giessenbachstr., 85741 Garching, Germany}

\author{Paula Jofre}
\affiliation{N\'ucleo de Astronom\'ia, Facultad de Ingenier\'ia y Ciencias, Universidad Diego Portales, Ej\'ercito 441, Santiago de Chile}

\author[0000-0002-5144-9233]{Andreas Just}
\affiliation{Astronomisches Rechen-Institut, Zentrum f\"ur Astronomie
der Universit\"at Heidelberg, M\"onchhofstr.\ 12--14, 69120 Heidelberg,
Germany}

\author[0000-0002-0129-0316]{Savita Mathur}
\affiliation{Instituto de Astrof\'{\i}sica de Canarias, La Laguna, Tenerife, Spain}
\affiliation{Dpto. de Astrof\'{\i}sica, Universidad de La Laguna, La Laguna, Tenerife, Spain}

\author{Andrea Miglio}
\affiliation{School of Physics and Astronomy, University of Birmingham, Edgbaston, Birmingham, B15 2TT, UK}
\affiliation{Stellar Astrophysics Centre (SAC), Department of Physics and Astronomy, Aarhus University,  DK-8000 Aarhus C, Denmark}

\author[0000-0002-5627-0355]{Ivan Minchev}
\affiliation{Leibniz-Institut f{\"u}r Astrophysik Potsdam (AIP), An der Sternwarte 16, 
14482 Potsdam, Germany}

\author{Giacomo Monari}
\affiliation{Leibniz-Institut f{\"u}r Astrophysik Potsdam (AIP), An der Sternwarte 16, 14482 Potsdam, Germany}
\affiliation{Observatoire astronomique de Strasbourg, Universit\'e de Strasbourg, CNRS, 
11 rue de l'Universit\'e, F-67000 Strasbourg, France }

\author[0000-0002-7547-1208]{Benoit Mosser} 
\affiliation{LESIA, Observatoire de Paris, PSL Research University, CNRS, Sorbonne Universit\'e, Universit\'e Paris Diderot,  92195 Meudon, France}

\author{Andreas Ritter}
\affiliation{The Laboratory for Space Research, Hong Kong University, 
Cyberport 4, Hong Kong SAR, PRC}

\author[0000-0002-9414-339X]{Thaise S. Rodrigues} 
\affiliation{INAF Astronomical Observatory of Padova, 36012 Asiago (VI), Italy} 

\author[0000-0002-0894-9187]{Ralf-Dieter Scholz}
\affiliation{Leibniz-Institut f{\"u}r Astrophysik Potsdam (AIP), An der Sternwarte 16, 14482 Potsdam, Germany}

\author[0000-0002-0920-809X]{Sanjib Sharma}
\affiliation{Sydney Institute for Astronomy, School of Physics, The University of 
Sydney, NSW 2006, Australia}

\author{Kseniia Sysoliatina} 
\affiliation{Astronomisches Rechen-Institut, Zentrum f\"ur Astronomie
der Universit\"at Heidelberg, M\"onchhofstr.\ 12--14, 69120 Heidelberg,
Germany}

\correspondingauthor{Matthias Steinmetz}
\email{msteinmetz@aip.de}
\collaboration{(The \rave\  collaboration)}

\begin{abstract}
{
The Radial Velocity Experiment (\rave) is a magnitude-limited ($9< I < 12$) spectroscopic 
survey of Galactic stars randomly selected in the southern hemisphere. The \rave\  
medium-resolution spectra ($R\sim7500$) cover the Ca-triplet region ($8410-8795$\,\AA). The 6th and final data release (DR6 or FDR) is based on 518\,387 observations of 451\,783 unique stars. \rave\ observations were taken between 12 April 2003 
and 4 April 2013. Here we present the genesis, setup and data reduction of \rave\ as well as wavelength-calibrated and flux-normalized spectra and error spectra for all observations in \rave\ DR6. Furthermore, we present derived spectral classification and radial velocities for the \rave\ targets, complemented by  cross matches with Gaia DR2 and other relevant catalogs. A comparison between internal error estimates, variances derived from stars with more than one observing epoch and a comparison with radial velocities of Gaia DR2 reveals consistently that 68\%\ of the objects have a velocity accuracy better than $1.4\,\kms$, while 95\%\ of the objects have radial velocities better than $4.0\,\kms$. Stellar atmospheric parameters, abundances and distances are presented in subsequent publication.  The data can be accessed via the \rave\  Web 
site\footnote{{\tt http://rave-survey.org}} or the Vizier database. }
\end{abstract}


\section{Introduction}

Deciphering the structure and formation history of the Galaxy provides important clues 
for understanding galaxy formation in a broader context. Wide field spectroscopic surveys 
play a particularly important role in the analysis of the Milky Way: Spectroscopy 
enables a measure of a star's radial velocity (RV), one of the six-dimensional coordinates
of position and velocity, which in turn allows us 
to study the details of Galactic dynamics. Spectroscopy also permits a measure of the 
abundances of chemical elements in a star's atmosphere, which holds important clues to 
the star's initial chemical composition and the subsequent metal enrichment of the 
interstellar medium traced by stars of different ages and metallicities
\citep[see, e.g.,][]{freeman2002,hawthorn2016}. 
However, despite the importance of stellar spectroscopy for Galactic dynamics and Galactic archaeology, 
the data situation in the early 2000s was far from satisfactory. RVs were listed for 
some 50,000 stars in the databases of the Centre de Donn\'{e}es astronomiques de 
Strasbourg (CDS), an astonishingly small number compared to the approximately one 
million spectra available for galaxy redshifts listed at that time. Furthermore, 
these RVs and their underlying spectra comprised a very heterogeneous sample in terms of selection, resolution, 
epoch or signal-to-noise ratio (\SNR). The situation changed somewhat with the advent 
of the Geneva Copenhagen survey \citep[CGS,][]{nordstrom2004}, which provided radial 
velocities, effective temperatures, and metalicities for a homogeneous sample of 
14,139 stars. However, this sample covered only a sphere of about 100 pc radius 
around the Sun (the so-called Hipparcos sphere).

The RAdial Velocity Experiment (\rave)  was originally set up as a pilot survey using 
the existing 6dF multi-object spectrograph at AAO's UK Schmidt telescope (UKST) to observe 
about 100,000 stars in $\sim 180$ nights of unscheduled bright time during the years 
2003-2005 \citep{steinmetz2003}. Spectra were to be taken covering the IR Ca triplet
region also employed by the Gaia RVS system \citep[see][]{recio-blanco2016}. 
Motivated also by the astrometric 
satellite mission concepts DIVA \citep[\emph{Deutsches Interferometer f{\"u}r Vielkanalphotometrie und Astrometrie, }][]{DIVA} and FAME \citep[Full-sky Astrometric Mapping Explorer, ][]{FAME}, 
this pilot survey was 
intended as a path finder for a considerably larger campaign targeting up to 40 million 
targets using a new Echidna-based multi-object spectrograph for the UKST, thus providing 
a vast kinematic database three orders of magnitude larger than any other survey planned 
in this period. While the DIVA and FAME missions were terminated in 2004, the results of 
the \rave\  pilot survey were very encouraging. In particular, in 
addition to radial velocities, the determination of relevant information on stellar 
atmospheric parameters and potentially even abundance ratios appeared feasible. 
Consequently \rave\  was continued for, eventually, a full ten year period, providing one 
of the largest data bases for stellar parameters and radial velocities. Meanwhile a 
series of five data releases (DRs) with an increasing number of targets and increasingly refined 
data products have been released: DR1 \citep{steinmetz2006} provided radial velocities 
derived from 25,274 spectra; DR2 \citep{zwitter2008} radial velocities and atmospheric 
parameters derived from 51,829 spectra; DR3 \citep{siebert2011} the full pilot survey 
with 83,072 spectra; DR4 \citep{kordopatis2013} employed a new and much more refined 
pipeline for stellar parameter determination, and provided radial velocities and stellar 
parameters based on 482,430 spectra; DR5 \citep{kunder2017} provided a new and enhanced 
calibration of the derived stellar parameters, included a new calibration of
giant stars based on information from the asteroseismic K2 mission, and linked \rave\  targets to 
the Tycho-Gaia astrometric solution of Gaia DR1 \citep[TGAS:][]{lindegren2016}. The 
\rave\  data releases were complemented by value added catalogs, including 
spectro-photometric distances 
\citep{breddels2010,zwitter2010,burnett2011,binney2014,mcmillan2018}, 
chemical abundances \citep{boeche2011,casey2017}, and automated spectral classification 
\citep{matijevic2012} as well as catalogs of active 
stars \citep{zerjal2013,zerjal2017} 
and of candidates for very metal-poor stars \citep{matijevic2017}. Furthermore, \rave\  
has meanwhile been complemented by surveys of similar or even larger size at lower (e.g., 
SEGUE \citep{yanny2009} and LAMOST \citep{LAMOST})  and higher spectral 
resolution (e.g., APOGEE \citep{APOGEE}, GALAH \citep{GALAH}, and 
Gaia-ESO \citep{gilmore2012}). {For a recent review on abundances derived from large spectroscopic surveys we refer to \cite{jofre2019}.}

The pair of this paper (DR6-1) and its accompanying paper \citep[DR6-2][]{steinmetz2020b} is the 6th and last publication in the series of \rave\  data release papers. DR6-1 will focus on the spectra taken and  is accompanied by a data base of wavelength calibrated and flux normalized 
spectra for 518\,387 observations of 451\,783 unique stars. DR6-2 provides a new set of stellar parameters employing parallax information 
from Gaia DR2 \citep{GaiaDR2}, a robust $\alphafe$ ratio, and 
individual [Fe/H], [Al/H] and [Ni/H] ratios. 

DR6-1 is structured as follows: 
in Section 2 we give an overview of the survey facility and performance. We outline the 
data reduction and provide direct references to sections of previous papers where the 
interested reader can find further details. Section 3 presents the spectra in the 
\rave\  spectral catalog, which we are releasing here for the first time, and the 
reduction procedure of the \rave\  raw data. Section 4 presents the automated classification of \rave\ spectra. Section 5 is devoted to the derivation of 
radial velocities.  \rave\ data validation including a 
comparison of \rave\  radial velocities with Gaia DR2 data is done in Section 6. Section 7 presents the \rave\ spectral DR6 catalog, radial velocities, classification, and crossmatch with other relevant catalog data. Finally, Section 7 gives 
a summary, draws some conclusions, and provides an outlook.

\section{Survey Description}\label{sec:SD}

Most of the technical specification and description of the Survey performance 
in terms of observational setup, procedure, and data reduction are outlined in the 
DR1-DR5 data release papers. Since this paper describes the final data release, we give 
an overview of the basic survey procedures and provide references to the sections in 
previous papers where the interested reader can find further details.

\subsection{Survey Facility}
\rave\  observations were performed at the 1.23m UK Schmidt telescope at Siding 
Spring in Australia using the 6dF multi-object spectrograph \citep{watson2000}, 
featuring a $5.7\degr$ field of view. 6dF consisted of an off-telescope robotic fiber 
positioner, two fiber field plates of 150 fibers each (three as of February 2009), and 
a bench-mounted spectrograph, mounted on the floor of the telescope dome. The 
spectrograph was fed from the UKST when one of the field plates was mounted to the 
telescope. Each fiber had a diameter corresponding to $6.7\arcsec$ on the sky and 
could be placed with an accuracy of $0.7\arcsec$ within the $\sim 6\degr$ diameter field. The 
spectrograph used a volume phase holographic (VPH) transmission grating 
of medium dispersing power; this 1700 lines mm$^{-1}$ grating was tuned for high 
efficiency in the I-band. This setup corresponded to an average resolving power of 
$R\approx 7500$ over the calcium triplet region at $8410-8795\,$\AA. The wavelength 
region covered by \rave\  is thus very similar to that probed by the Gaia RVS 
instrument \citep{Cropper2018} at somewhat lower average resolution 
($R_\mathrm{RVS}=11,500$). The CCD used in the 6dF instrument was a 
Marconi (EEV) CCD47-BI detector that features $13\,\mu$m pixels in a $1056\times 1027$ 
array. It had a quantum efficiency of 30-40\% in the wavelength region adopted by 
\rave. For further details we refer to DR1, Section 2.1.

Each field plate featured $\approx 150$ fibers deployed from a ring around the 
periphery of the $5.7\degr$ field. Each fiber could nominally reach 10\% past the field 
center and was constrained to an angle of $\pm 14\degr$, resulting in subtle allocation 
biases \citep[][see also Figure 3 in DR1]{Miszalski2006}. The actual allocatable fiber 
numbers typically varied between 100 and 120 (but could be as low as 80 immediately 
before fiber bundles were refurbished). The most common problem for fiber 
unavailability were fiber breakages while parking fibers. Other problems include 
deterioration of fiber throughput or problems for the robot picking up fiber buttons. 

Prior to configuring a field, each target was drawn from the input catalog (see 
Section \ref{subsec:input}) 
based on priorities given within the input target list. The targets were then
manually checked for contamination, double star proximity and variability by downloading 
thumbnail images from the Supercosmos 
Sky Survey \citep[SSS, ][]{Hambly2001b} that are large enough to cover the fiber’s field of view. 
Contaminated stars 
were replaced until a clean, homogeneous field was achieved. Each candidate was then 
allocated to a given fiber using a sophisticated field configuration algorithm based 
on that developed for the 2dF spectrograph \citep{Lewis2002}.  The field configuration 
algorithm accepts an user-supplied input 
catalog and configures fibers based on priorities given within the input target list.

Configuring a full field plate typically took about an hour, a relevant boundary 
condition for setting the typical exposure time (and magnitude of the targets). At a 
magnitude of $I=10 - 11$, the exposure time to reach a $\SNR > 40$, the target density of 
objects of that magnitude at the Galactic poles within a $5.7\degr$ field, and the 
configuration time fitted neatly together to give a sensible exposure time of $\approx 1\,h$.

After the conclusion of the \rave\  survey in April 2013, the 6dF facility 
(spectrograph, robot, and positioner) was decommissioned and taken out of operation.

\subsection{Survey design and input catalog}\label{subsec:input}

\rave\  was designed to be a magnitude limited spectroscopic survey that avoids any 
kinematic biases in the target selection. The magnitude range probed corresponds to 
$9<I<12$, where $I$ is Cousins $I$. No color selection was performed (see however the discussion below for the added fields at low Galactic latitudes). The wavelength range of \rave\  of $8410-8795$\,\AA\ overlaps with the photometric 
Cousins $I$ band.

When the \rave\  survey started preparation for the first years of operation in 2002, 
neither the 2MASS \citep{2MASS} nor the DENIS \citep{DENIS} catalog was available. 
Therefore, \rave\  targets 
stars were drawn from the Tycho-2 catalog \citep{Hog2000} and from SSS. 
For the Tycho I-band, the magnitudes were estimated using  the transformation formulae from \cite{Perryman1997} and \cite{Bessell1979}. 
The photographic $I_\mathrm{IVN}$ magnitudes in SSS are directly equivalent to 
Cousins $I$ \citep{Blair1982} and no further transformation was applied. 

Stars between $11<I<12$ were exclusively drawn from SSS. Stars with $9<I<11$ originate 
predominantly in Tycho-2, but SSS stars that do not appear within 6.7\arcsec\ 
(corresponding to the size of a fiber on the sky) of a Tycho star are included as well. 
We also did not include stars in Tycho 2 or SSS that were within 6.7\arcsec\ of another 
Tycho-2/SSS star to exclude possible contamination by unresolved multiple sources .  
For the same reason, i.e., to avoid unresolved multiple sources within a single fiber, the 
initial input catalog was limited to fields at Galactic latitudes of $b>25\degr$, but 
for observing efficiency reasons (available sky regions observable with UKST for given 
observing epoch), fields with $15\degr <|b|\leq25\deg$ were subsequently added for 
all Galactic longitudes ($\ell$). The field centers of the first input catalog are 6\degr\ apart.

The early input catalog of the first 2 years of operation thus contained about 
300,000 stars of which about half the sample originated from Tycho 2, the other 
half from SSS. The first three data releases of \rave\  (DR1, DR2, and DR3) are 
entirely based on this input catalog. An {\it a posteriori\/} comparison with DENIS DR3 \citep{DENIS_DR3} revealed that, owing to saturation effects for $I<13$, a $\approx 1$~mag offset 
between DENIS and SSS at $I\approx 11$ emerged. As a consequence, while \rave\  
comprises a kinematically unbiased sample, the early input catalog exhibits some 
color biases (see discussion in DR1 Section 2.3).

A new and more refined input catalog was brought into use in March 2006.  The main 
sample has $|b| > 25$, and uses DENIS DR3 cross-matched with 2MASS to define targets to 
$I$ = 12 with a default of four pointings on each field center -- two bright 
and two faint. The field centers are now 5 degrees apart to ensure some overlap between 
adjacent fields. With the new input catalog, also an attempt was made to more 
carefully extend the input catalog to lower Galactic latitudes, i.e., to include more 
of the Galactic disk towards the Galactic anti-rotation direction ($225\degr < \ell < 
315\degr$, $5\degr <|b|<25\degr$). A mild color cut of $J-K >0.5$ was used in this 
region to avoid observing young stars, as the weak Paschen lines 
in the CaT region mean that radial velocities and, in particular, stellar parameters,  
can be only poorly determined (see Section \ref{subsec:samplespec} and Figure 
\ref{fig:sample_spec} below).  

In the post-2010 operations, the input catalog was further extended, again for reasons 
of observing efficiency, to lower Galactic 
latitudes and thus closer to the Galactic mid plane, so that 
reddening had to be taken into account. The aforementioned color cut of  $J-K >0.5$ 
is capable of rejecting young foreground stars provided that  $E(B-V) < 0.35$\,mag. Thus, low latitude fields  ($10\degr<b<25\degr$) are included for 
$315\degr< \ell <330\degr$ and $\ell<225\degr$, and, analogously (for
$-25\degr < b < -10\degr$), fields with Galactic longitudes of $\ell<225\degr$, $\ell>315\degr$ and $\ell<30\degr$, respectively.

The fields observed at $|b|<25\degr$ of the old input catalog (no color cut) compared 
to the new input catalog (with color cut $J-K>0.5$) can be easily identified by having an 
observing date $\leq 20060312$. The stars from these earlier fields are excluded from the selection function,  
as discussed in Section \ref{subsec:SF}.

Finally, and again for observing efficiency reasons, targets for $0\degr\leq\delta \leq 
5\degr$ and $0^\mathrm{h}\leq\alpha\leq 6^\mathrm{h}$, 
$7^\mathrm{h}30^\mathrm{m}\leq\alpha\leq 17^\mathrm{h}$, 
and $19^\mathrm{h}30^\mathrm{m}\leq\alpha\leq 24^\mathrm{h}$ were needed. However, no DENIS counterpart was available  for 
targets north of $\delta=2\degr$. Targets were therefore  
defined from 2MASS, with their estimated $I$ magnitudes derived from 2MASS $J$ and $K$, following  
equation~1 in DR4.

In addition to  the survey fields described above,  a number of targeted observations were 
performed that focused on a selection of open and globular clusters. These fields were acquired to 
allow independent checks on the \rave\  stellar parameters and their errors (for 
details, see DR5 Section 7.1).

For details regarding the input catalogs we refer to DR1 Section 2 and to DR4 
Section 2, respectively.

\subsection{Observing procedure}\label{subsec:obs_procedure}

Observations for \rave\  followed a sequence of target field exposures, arc and flat. 
Ne, Rb, and Hg-Cd calibration exposures were obtained for each field, together with a 
quartz flat field for spectrum extraction in the data reduction. Typically, one 10s 
RbNe arc exposure and five 15s fiber flats (quartz halogen) were taken before and after 
each field exposure series. The field exposure series themselves consisted of five 
consecutive exposures (see below), allowing adequate \SNR\  to be obtained in the 
summed spectra, while minimizing the risk of saturation from particularly bright stars. 
In the case of poor conditions or low sky transparency, additional exposures were made. 

Several target fibers were reserved in order to monitor the sky for background
subtraction. Each of the \rave\  target frames contained
spectra of at least 10 sky samples, obtained using dedicated sky
fibers. These were combined and scaled in the reduction process
for sky subtraction.

We used the two field plates (three field plates starting in 2009) on an 
alternating basis, i.e., fibers from one field plate were configured while we 
observed with the other field plate. So fibers from a given field plate
were mounted to the spectrograph slit prior to the observation of each field. To do 
this the cover of the spectrograph needed to be removed, so its temperature might 
change abruptly. Because of the associated thermal stress we took the flatfields
and neon arc lamp exposures immediately after the set of scientific exposures, i.e., 
at a time when the spectrograph was largely thermally stabilized.

Taking account of the physical
transportation and exchange of the field plates, the slew time
for the telescope, field acquisition, etc., an experienced observer was
able to accumulate acceptable data for up to eight \rave\  fields on
a midwinter{'}s night at the latitude of Siding Spring Observatory.

\subsection{Major changes in the performance of the \rave\  survey}

\rave\  observations span a period of ten years. Based on the lessons learned with 
early data releases, in particular DR1 and DR2, a number of procedural optimizations 
were introduced. Furthermore, maintenance and refurbishments of the telescope and the 
facility resulted in a few modifications.  We summarize the most 
relevant ones in the following:
\begin{itemize}
    \item {The red color of early selected targets \citep[DR1,][]{steinmetz2006} and a predicted
 low efficiency of the spectrograph in its 2nd order did not call initially
 for a blue-light blocking filter. A Schott OG530 blue-light blocking filter was
 however inserted in the collimated beam of the spectrograph on 2 April
 2004, to fully suppress the contamination visible expecially on warmer
 targets.}
    This allows for an unambiguous placement of the continuum level and so permits the 
    derivation of values of stellar parameters, in addition to the radial velocity 
    (DR2). 
    \item \rave\  observations were initially limited to 7 nights of bright time per 
    lunation owing to the then ongoing 6df Galaxy Redshift Survey 
    \cite[6dFGRS][]{Jones2009}. With the conclusion of 6dFGRS on 31 July 2005, \rave\  
    proceeded through the end of 2012 at an observing rate between 20 and 25 nights per 
    lunation.
    \item On 13 March 2006, the new DENIS+2MASS based input catalog was introduced 
    (see Section \ref{subsec:input} and DR4 Section 2).
    \item On 29 March 2006, the number of fiber flats was increased from 1 to 5. 
    \item \rave\  observations initially consisted of 5 exposures  
    of 600\,s. Since the beginning of 2007, \rave\  targets were segregated into 
    four magnitude bins (bin$_1$: $8\leq I\le 10$, bin$_2$: $10\leq I < 10.8$, bin$_3$: 
    $10.8 \leq I < 11.3$, and bin$_4$: $ 11.3\leq I < 12$) in order to maximize 
    observing efficiency and to avoid cross-talk contamination of fibers on faint 
    sources adjacent to fibers targeting bright objects. Exposure times corresponded to 
    $5\times 600$\,s  for bin$_1$ and bin$_2$, $5\times 900$\,s for bin$_3$, and 
    $5\times 1200$\,s for bin$_4$. 
    \item Observations were paused between 4 June 2007 and 11 June 2007 for service on 
    the 6dF robot and between 26 June 2007 and 6 August 2007 for asbestos removal work at 
    the UKST.
    \item In February 2009 a third field plate was introduced and subsequently the 
    original two field plates were fully refurbished with new fiber bundles. After this 
    procedure, each observing night started with two fully configured field plates, thus 
    considerably increasing the survey speed.
    \item The Wambelong bushfire at Siding Spring in early 2013 forced observations to 
    be suspended between 13 January 2013 and 1 April 2013. 
\end{itemize}

\subsection{Data reduction}\label{subsec:reduction}

The data reduction of \rave\  follows the sequence of the following pipeline:
\begin{enumerate}
    \item quality control of the acquired data on site with the RAVEdr software 
    package (Section \ref{subsec:spectra}).
    \item reduction of the spectra (Section \ref{subsec:spectra}).
    \item spectral classification (Section \ref{subsec:Class}).
    \item determination of (heliocentric) radial velocities with \texttt{SPARV} 
    (\emph{`Spectral Parameter 
And Radial Velocity'}, Section \ref{sec:SPARV}).
    \item determination of atmospheric parameters with \texttt{MADERA} (\emph{`MAtisse and DEgas used in RAve'}, (paper DR6-2, Section 3.1, see also \citealt{kordopatis2013}).
    \item determination of the effective temperature using additional photometric 
    information (\emph{InfraRed Flux Method} (\texttt{IRFM}), DR6-2, Section 3.2, see also \citealt{kunder2017}).
    \item determination of atmospheric parameters combining \rave\  
    spectroscopic information with additional photometry and Gaia DR2 parallax priors 
    using \texttt{BDASP} (\emph{Bayesian Distances Ages and Stellar Parameters}, DR6-2 Section 3.3, see also \citealt{mcmillan2018}).
    \item recalibration of the stellar parameters for giant stars based on K2 asteroseismic information (DR6-2, Section 3.4, see also \citealt{valentini2017}).
    \item determination of the abundance of iron group elements and an overall $\alphafe$ ratio with the pipeline GAUGUIN 
    (DR6-2, Section 4).

\end{enumerate}

The output of these pipelines is accumulated in a PostgreSQL data base and accessible via the \rave\  website {\tt http://www.rave-survey.org} (Section \ref{sec:FDR} and DR6-2, Section 7).

\begin{deluxetable}{lrr}
\tablecaption{\label{ravestats}
Contents of \rave\  DR6.}
\tablehead{\colhead{in DR6}  & \colhead{N. of spectra}  & 
\colhead{N. of unique stars}}
\startdata
Observed targets & 518,387 & 451,783 \\
-- with \texttt{snr\_med\_sparv} $>20$ & 474,649 & 416,365\\
-- with \texttt{snr\_med\_sparv} $>40$ & 262,199 & 232,282\\
-- with \texttt{snr\_med\_sparv} $>60$ & 66,815 & 58,992\\
-- with \texttt{snr\_med\_sparv} $>80$ & 14,056 & 12,417\\
with 2MASS cross match & 518,300 & 451,706 \\
with Gaia DR2 cross match & 517,095 & 450,641 \\
\enddata
\end{deluxetable}

\begin{figure}
\begin{center}
\plotone{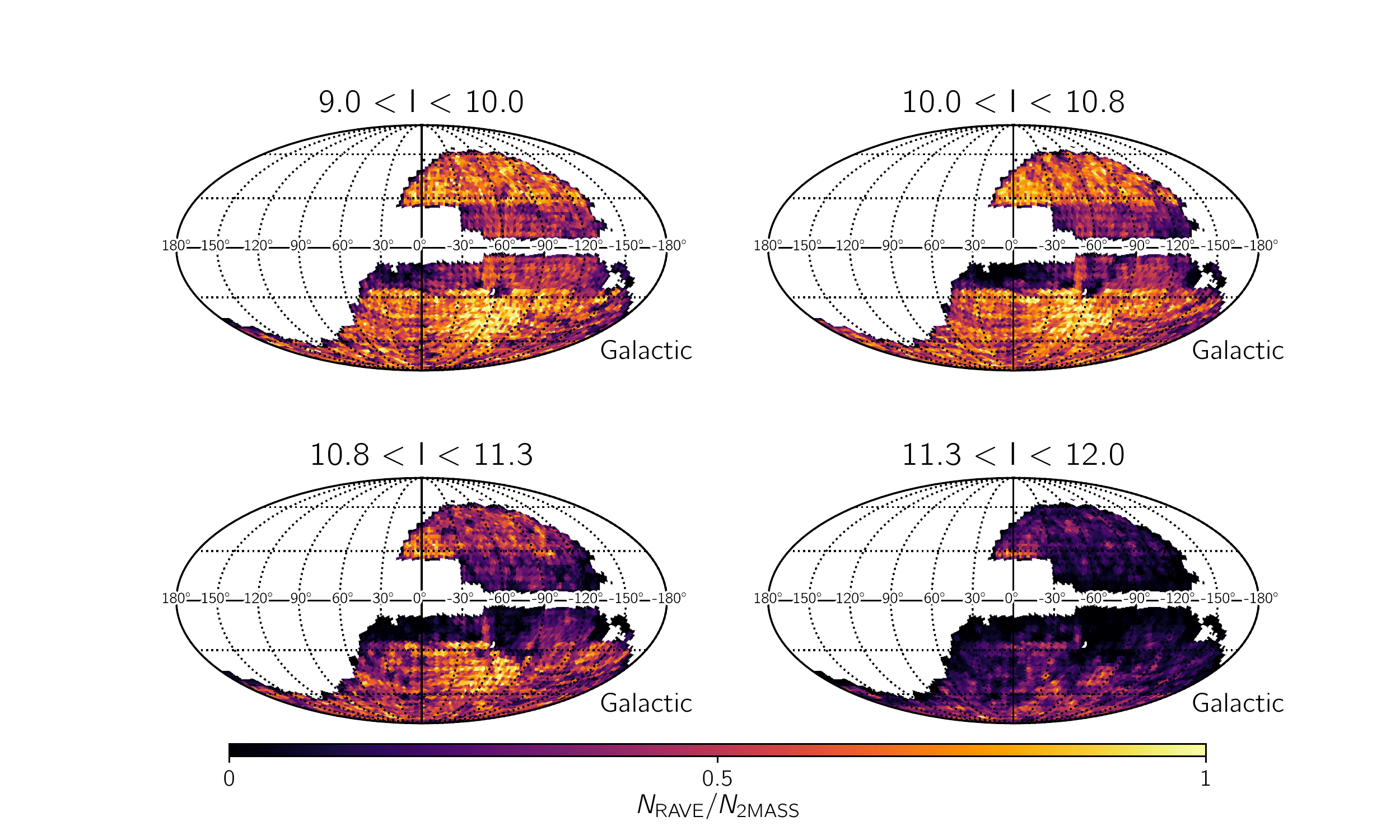}
\caption{\label{fig:rave_completeness}
Mollweide projection of Galactic coordinates of the completeness of
the stars for which \rave\ DR6 radial velocity measurements are
available for the core sample (see Section \ref{sec:Validation}). Each panel shows the 
completeness over a different magnitude range, where the HEALPix pixels are color-coded 
by the fractional completeness ($N_\mathrm{\rave}/N_\mathrm{2MASS}$).}.
\end{center}
\end{figure}

\subsection{Survey Selection Function}
\label{subsec:SF}
\begin{figure*}
\begin{center}
\plotone{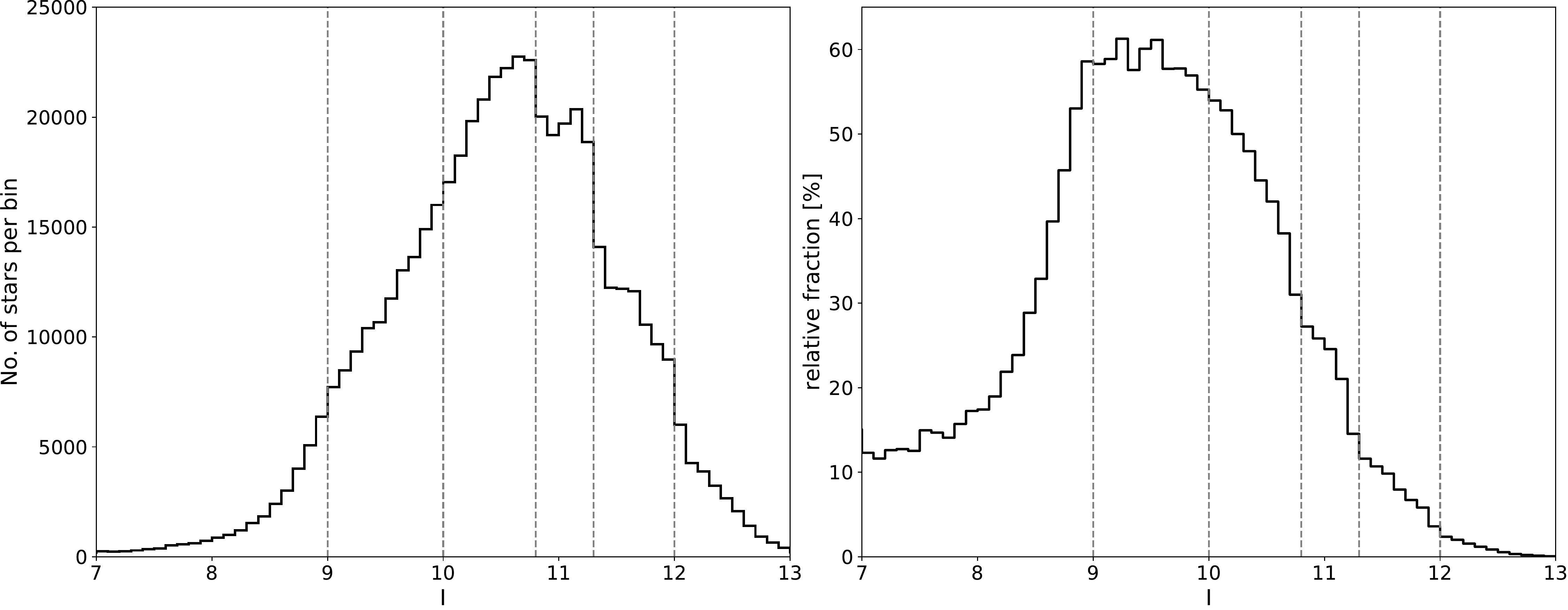}
\caption{\label{fig:rave_complete_cumm}Left: histogram of the number of spectra with derived radial velocities in the \rave\ footprint {per $I$ magnitude bin of 0.1} .  Magnitude bins used per field plate are indicated with dashed lines (see Section \ref{subsec:input}). Right: The completeness of RAVE
DR6 (stars with radial velocities) with respect to the completeness of 2MASS is shown as a function
of the $I$ magnitude.}
\end{center}
\end{figure*}

In order to draw robust conclusions from the data gathered via large spectroscopic 
surveys such as \rave, it is crucial to understand the relationship between the observed 
targets and their underlying population, known as the selection function. A 
comprehensive overview of the selection function of \rave\  is given in 
\citet{wojno2017}, which we summarize here. \rave\  targets were selected from a number 
of input catalogs. These targets were selected uniformly over the entire Southern 
hemisphere, with the exception of regions where a mild color-cut of 
($J - K > 0.5$ mag) was enforced (Section \ref{subsec:input}). 
Figure~\ref{fig:rave_completeness} shows the completeness fraction (number of \rave\ 
stars divided by number of 2MASS stars per area on the sky) for the observed $I$ magnitude 
bins. The $I$ magnitude is in principle available from catalogs as DENIS, however DENIS $I$ suffers from saturation effects for $I<10$. As in \citet{wojno2017}, we approximate the DENIS $I$ magnitude from 2MASS $J$ and $K_s$ via
\begin{equation}
    (I-J)=(J-K_s)+0.2\exp\frac{(J-K_s)-1.2}{0.2}+0.12
\end{equation}
{\citep[for the number of spectra based on measured APASS $i'$ magnitude (for \rave\ DR4), see Figure 11 of ][]{munari2014}}.
Figure \ref{fig:rave_complete_cumm} shows the completeness of stars with determined radial velocities over the overall \rave\ 
footprint.  As \cite{wojno2017} showed for \rave\  DR5 , we also find \rave\  DR6 to be 
kinematically unbiased.

\subsection{Repeats}\label{subsec:Repeats}

\rave\  was designed as a survey with its main focus on studies of Galactic dynamics 
and Galactic evolution. The primary design goals were therefore to have an unbiased 
input catalog and observing procedure, a wide coverage of the accessible sky, and a 
magnitude limited layout aiming at high completeness from the brighter to the fainter 
magnitudes. The technical boundaries (large multiplex and long configuration times 
combined with weather patterns and block-out periods around the full moon) made it 
difficult to account systematically for repeat observations following a fixed cadence, in particular without 
compromising  the aims of coverage and completeness. On the 
other hand, repeat observations are critical to check for coherency 
and repeatability of \rave\  data products, and to allow modellers to account for the 
effects of unresolved binaries.

\begin{figure}
\begin{center}
\plotone{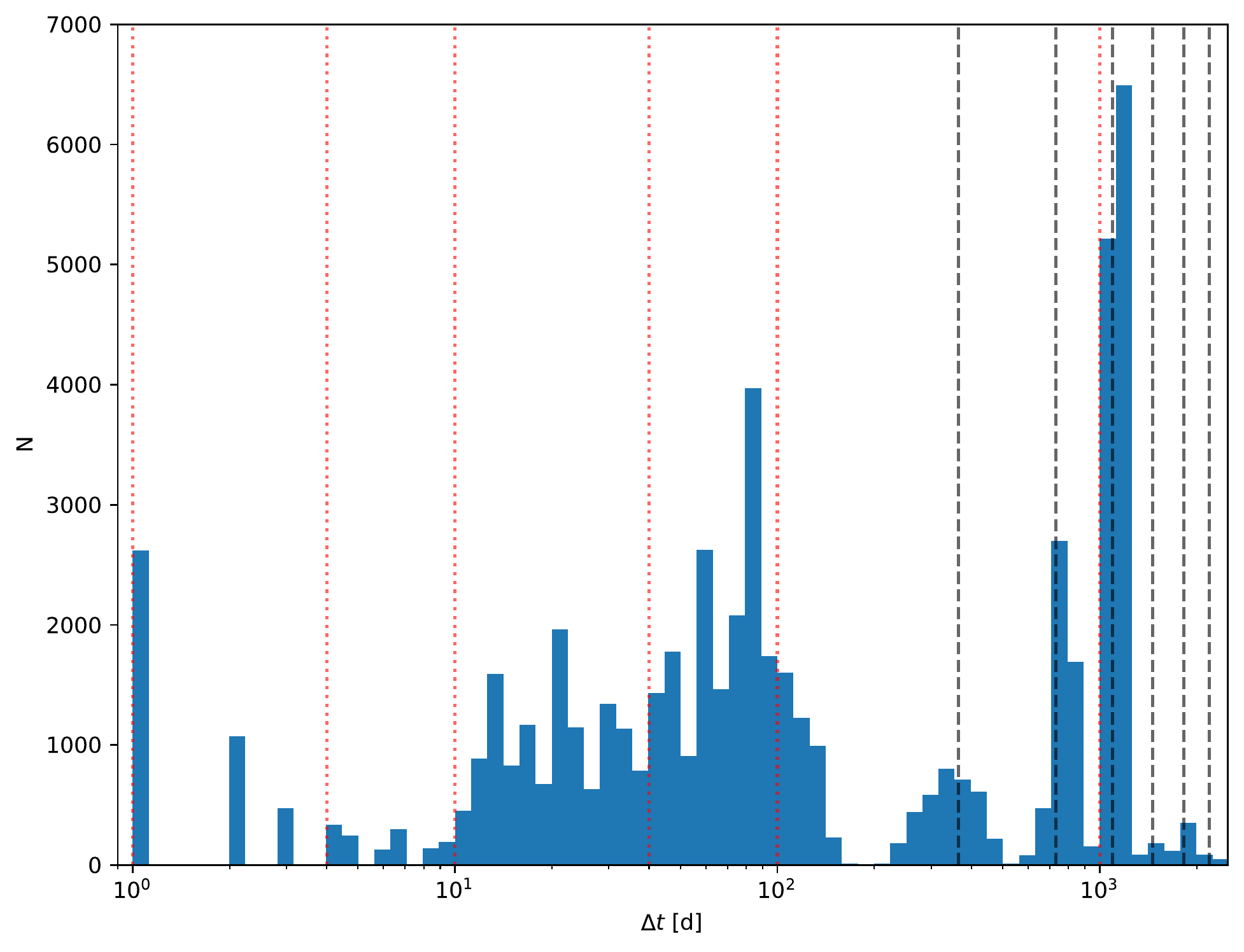}
\caption{\label{fig:rep_frequency}
Time interval between consecutive observations of stars with at least 4 repeat 
observations. The red dotted lines mark the guiding cadence of 1, 4, 10, 40, 100, and 1000 
days, the black dashed line multiples of 365 days.}
\end{center}
\end{figure}

\begin{figure}
\begin{center}
\plotone{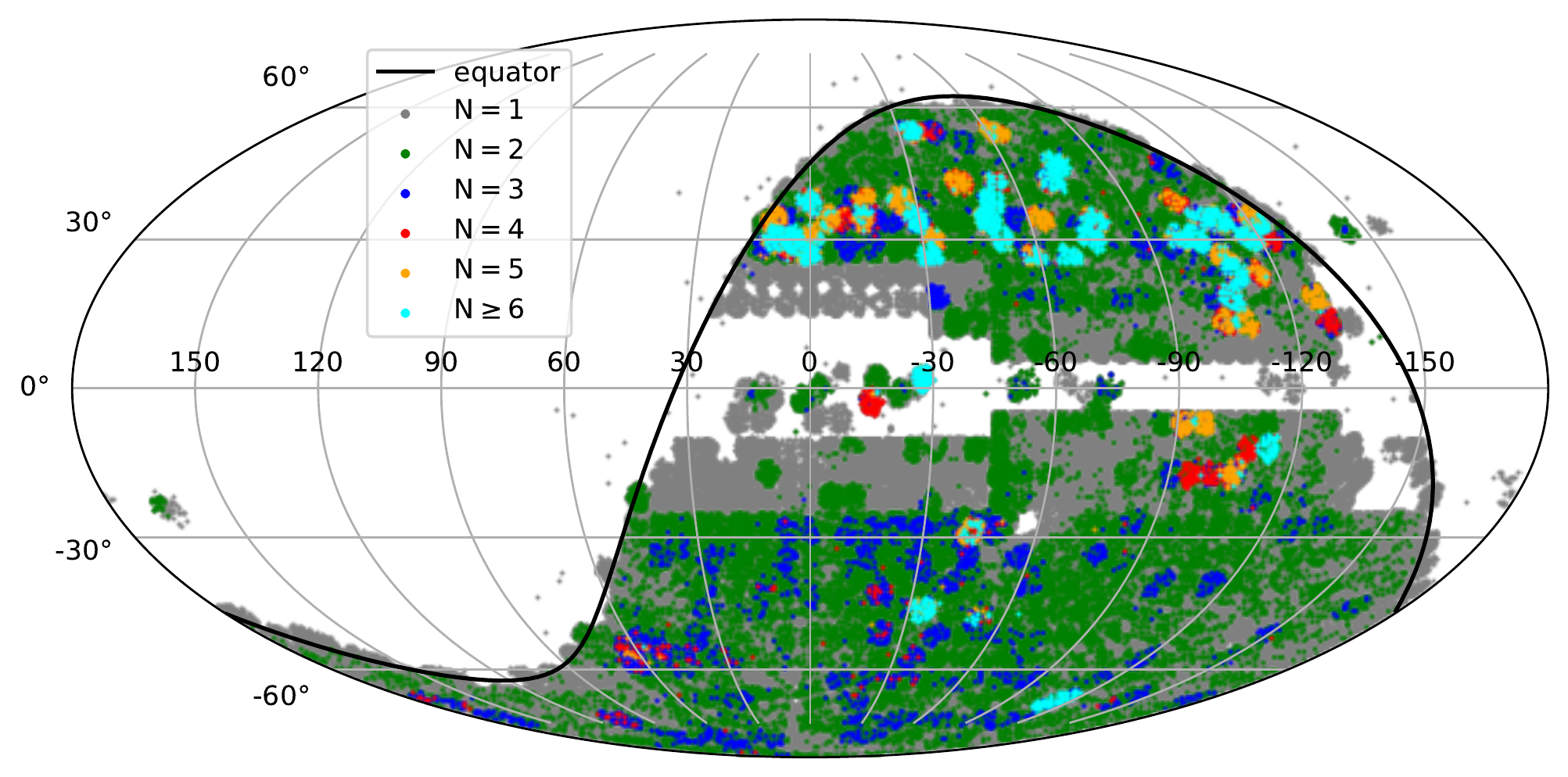}
\caption{\label{fig:rep_mollweide}
Mollweide projection of \rave\ fields color-coded by the number of revisits.}
\end{center}
\end{figure}

In order to measure, at least statistically, the effects of binarity, about 4000 stars 
were selected for a series of repeat observations in the 
observing semester 2009A (1 February 2009 - 31 July 2009). The aim was to roughly follow a logarithmic series with a cadence of 
separations of 1, 4, 10, 40, 100, and 1000 days. The repeat sequence was selected from 
the first observations of the new input catalog introduced on 12 March 2006, so the 
difference between the 2009 and the 2006 observations served to approximate the 1000 
day separation. Weather patterns, block out periods, and fiber availability, however, 
resulted in considerable dispersion and non-observations around the target cadence 
(Figure~\ref{fig:rep_frequency}). The repeat sequence mainly constitutes 
the subsets with 4, 5, 6, and more than 7 observations (see Table \ref{tab:repeats}), 
and can nicely be seen in the clustering at $-120\degr < l < 30\degr$ and $b>+30\degr$ 
(Figure~\ref{fig:rep_mollweide}).

In addition to  these systematic repeats, whole fields were repeated when they were marked as 
problematic in the post-observation quality review. Furthermore, individual stars could 
be re-targeted if no higher-priority (i.e. unobserved) targets were available in the 
fiber configuration process. These quality repeats and chance repeats make up most of 
the targets with 2 or 3 visits. Indeed, Figure \ref{fig:rep_mollweide} reveals that 
these targets are much more evenly spread over the \rave\  footprint, as expected. 
Finally, targeted observations of calibration fields, in particular open and globular 
clusters, also give rise to many repeat observations, also visible in the 
Mollweide projection of Figure~\ref{fig:rep_mollweide}.

\begin{deluxetable}{lr}
\tablecaption{\label{tab:repeats}
Stars with multiple observations in \rave\  DR6.} 
\tablehead{\colhead{}
 & \colhead{in DR6}}
\startdata
Stars with 1 visit  & 404,428 \\
Stars with 2 visits  & 39,340 \\
Stars with 3 visits  & 3,606 \\
Stars with 4 visits & 1,034 \\
Stars with 5 visits & 1,418 \\
Stars with 6 visits & 1,205 \\
Stars with $\ge$7 visits & 757 
\enddata
\end{deluxetable}

\section{\rave\  spectra}

\subsection{Spectra and their Reduction}\label{subsec:spectra}
All \rave\  spectra were reduced with a semi-automated pipeline based on a 
sequence of dedicated \textsc{Iraf} routines. The use of a pipeline approach ensures a 
proper uniformity of reductions, while the requirement of specific human-approved 
standardized checks increase their reliability. The pipeline is described in DR1 and 
DR2. Here we summarize its main features and report on experience gained over a decade 
of its use. We also add specific information that is relevant for legacy purposes and 
that is important for a reader who would like to understand the underlying 
systematics.

To account for the temperature sensitivity of the spectrograph (Section 
\ref{subsec:obs_procedure}), we adopted a policy where each set of scientific exposures 
was accompanied by a flat-field and an arc-line exposure, both usually done immediately 
after the scientific ones. Flat-field exposures were used to establish position, width 
and shape of spectral tracings in other exposures, {to normalize relative fiber throughput} and to filter-out interference 
fringing, which can be quite prominent when using a thinned backside illuminated CCD 
detector at the \rave\  wavelength range. Typically, fringing jumps that reach up to 
20\%\ of the flux are damped to $\sim 1$\%\ with this approach, but note that 
techniques of Gaussian filtering would be required to reach a better continuum 
normalization on scales of $\sim 20$~\AA, a typical fringe width. A small fraction of 
\rave\  spectra suffer from internal reflection of light within the spectrograph, which 
causes an emission ghost located blue-ward of the Ca~II 8498~\AA\ line. This is an 
additive feature that can be removed through careful spectral normalization, though its 
width of $\sim 23$~\AA\ makes this a challenging task. Such spectra have a problematic 
continuum, so they are flagged with `c' (Table \ref{tab:Class}) in the final database.

Extracted one-dimensional spectra of object, flat-field, and arc-lamp exposures need 
to be corrected for fiber cross-talk and for scattered light contributions from all 
fibers with flux levels above a set threshold. Fiber cross-talk is removed iteratively, 
assuming a Gaussian shape of fiber illumination in the direction perpendicular to the 
tracing direction. In the end, we estimate that in the final spectrum the contribution 
of stars to adjacent fibers does not exceed 0.001 of their flux. About 13\%\ of the
incoming light is scattered in the spectrograph, with the exact amount decided manually 
by minimizing the flux in the gaps between the three fiber sub-bundles (fibers 50-51 
and 100-101) and by analyzing flux levels in sky fibers that have very low continuum 
levels or, in the case of cirrus clouds and moonlight, should have positive fluxes 
compatible with the strength of the Solar spectrum in lines of the calcium IR triplet. 
In our model the scattered light from each point in the focal plane is scattered over 
an axially symmetric Gaussian with a FWHM of ~$\sim 200$ CCD pixels. 

A neon arc lamp exposure is used to wavelength calibrate the spectra. The lamp 
includes 9 emission lines in the \rave\ spectral range that are strong enough for this 
purpose. Table \ref{tab:arclamp} reports their adopted wavelengths. Note that each 
spectrum includes 1031 pixels spanning a wavelength range of $384.6 \pm 1.7$~\AA, but 
its central wavelength varies in a parabolic manner from $\sim 8595$~\AA\ at the edges of 
the field plate to $\sim 8604$~\AA\ at its center (see Figure 3 in DR2). Consequently, 
the Ar~I 8408 \AA\ arc line is missing in fibers near the center of the field plate 
while at the edges of the field plate extrapolation has to be used to wavelength 
calibrate the reddest 15~\AA\ of the wavelength range. All observed spectra have been 
fit using 5 or more arc lines and 91\%\ have been fit using $N = 8$ arc lines. A single 
cubic spline with $\mathrm{df}=4$ free coefficients is used to convert pixel units into 
wavelengths. Per line it achieves a median difference between the fitted and assumed 
wavelength of 0.015~\AA\ (for 99\%\ of the spectra, this difference is smaller than 
0.072~\AA), which at 8600~\AA\ corresponds to $0.52$~$\kms$. This implies a typical error 
in the derived radial velocity of $\sim 0.52 / \sqrt{N-\mathrm{df}}$~$\kms = 0.26$~$\kms$. 
So the radial velocity accuracy is  mostly limited by temperature shifts in the 
spectrograph, by the achieved \SNR, and by the accuracy of flat-fielding, and not by 
uncertainties in the wavelength calibration.  

\begin{deluxetable}{ll}
\tablecaption{\label{tab:arclamp}
Wavelengths of arc lines.} 
\tablehead{\colhead{Element} & \colhead{Wavelength [\AA]}}
\startdata
\ion{Ar}{1} & 8408.2096 \\
\ion{Ar}{1} & 8424.6475 \\
\ion{Ne}{1} & 8495.3598 \\ 
\ion{Ar}{1} & 8521.4422 \\ 
\ion{Ne}{1} & 8591.2583 \\ 
\ion{Ne}{1} & 8634.647 \\ 
\ion{Ne}{1} & 8654.3831 \\ 
\ion{Ne}{1} & 8780.621 \\ 
\ion{Ne}{1} & 8783.7533 
\enddata
\end{deluxetable}

In 90\%\ of the cases the spectra of sky fibers are dominated by sky emission lines and 
have negligible continuum levels, while the rest include a detectable level of 
moonlight, usually scattered on thin cirrus clouds. The main contributor to sky spectra 
are therefore airglow emission lines belonging to three series: OH transitions 6-2 at 
wavelengths shortward of 8651~\AA, OH transitions 7-3 at wavelengths longward of 
8758~\AA, and O$_2$ bands between 8610~\AA\ and 8710~\AA\ (see DR5 for details). Both 
airglow and cirrus scattering can vary on shorter timescales and smaller spatial scales 
than a typical $\sim 50$-minute sequence of scientific exposures over a 5.7\degr\ field 
of view. Therefore, we assume a complete scrambling and use a scaled median of sky 
fibers as the background model. The user should be aware that in rare cases this may 
not be true, as both airglow line intensity and cirrus cloud scattering may depend on 
fiber position. In such cases objects with adjacent positions on the sky should show 
similar levels of sky residuals. We made sure that to the best of our knowledge the sky 
fibres were positioned on ``empty'' regions of the sky. We use two additional checks to 
avoid sky over-subtraction due to contamination by unknown sources: first, the person 
responsible is asked to visually approve all sky spectra to be used in the background 
calculation and second, the use of a scaled median rejects any remaining outliers. 

Airglow emission lines have fixed wavelengths, so they have been used for the calculation 
of the radial velocity zero point and for the correction of temperature fluctuations in the 
spectrograph. Their signal is much cleaner in the sky fibers than in the stellar spectra. The 
zero-point correction is obtained from a weighted sum measurement of sky and stellar 
spectra, with the former having a 10-times higher weight {and typically amounts to $\approx -0.5\pm 1$\kms}.   

The final stages of reduction include sky subtraction and shifting of the stellar 
spectrum to the inertial frame of the Solar system barycenter. Note that at all stages 
of the reduction the wavelength bin corresponds to one pixel in the dispersion 
direction. This simplifies the recognition and treatment of discrete features, like 
cosmetic defects of the CCD or cosmic ray hits. A sequence of $\sim 5$ 10-minute 
exposures of each object is median-combined, thus rejecting most of the cosmic ray hits 
(except for the rare hits in the flat-field exposure). 

As a final check, the pipeline makes two graphs that are visually inspected and 
stored: the first plot compares measured average fluxes in individual fibers with the 
ones expected from the available ground-based photometry. Since fiber throughput varies 
over time and position, we advise the user to use normalized 
spectra only. Still, a comparison of fluxes should show a clear correspondence with 
stellar magnitudes and low or negligible fluxes in sky and parked fibers. This helps to 
avoid any book-keeping errors, which are always possible when observing hundreds of 
thousands of stars over thousands of nights.  Finally, the responsible person is shown 
a collection of final spectral tracings of a given field: the idea is to check that
the results seem reasonable, but a visual check of every spectrum is not feasible  for such 
a large survey. {The design of the spectrograph does not allow to derive accurate absolute fluxes, so we provide spectra with a normalized continuum. These are derived with an iterative  low-order polynomial fitting and with asymmetric rejection limits. We used a second-order spline function with the upper rejection limit set to 2 residual standard deviations and lower limit to 1.3. Note, however, that any comparison of observed and synthetic spectra requires that {\it both} spectra are first normalized {\it again} using the same normalization parameters. 
}

In summary, \rave\  uses a dedicated data reduction pipeline, which has been tuned by 
our experience gained over a decade of its use. Our insistence on specific human 
approved checks and adjustments increases the reliability of its results. Note that this 
is different from surveys that use general-purpose instruments and often use or at least 
start with the instrument supplied pipeline. On the 
other hand, the HERMES spectrograph is a general purpose instrument at the AAO, but its 
extensive use by the GALAH survey makes the survey's dedicated pipeline \citep{kos2017} 
increasingly popular also for general users as well. Finally, we note that all these 
pipelines use the standard reverse-modelling approach. It seems that with photonic 
combs an alternative forward modelling approach is possible, which convolves a list of 
spectral templates with assumed values of stellar parameters with known aberrations in 
the spectrograph to produce a fit to the original CCD image. This approach may yield 
much better results in the future \citep{kos2018}.

\begin{figure*}
\begin{center}
\epsscale{1.15}
\plotone{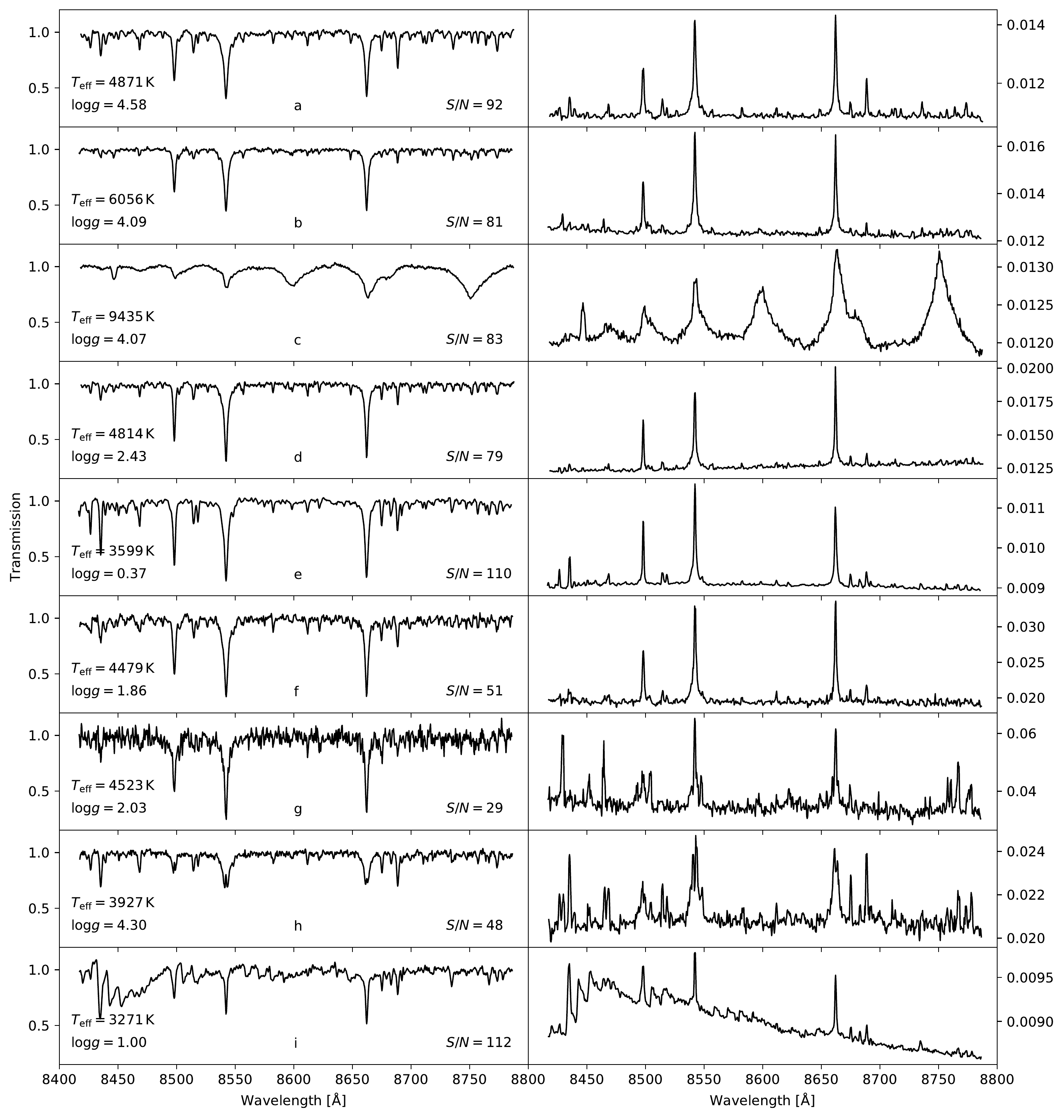}

\caption{\label{fig:sample_spec}
Typical spectra (left) and error spectra (right) for some typical objects in the \rave\ 
data base. Stellar parameters are derived using the \texttt{BDASP} pipeline (see DR6-2, Section 3.3). 
From top to bottom: a) a high \SNR\ cool dwarf; b) a high \SNR\ warm dwarf; c) a 
high \SNR\ hot dwarf; d) a high \SNR\ red clump star; e) a high \SNR\ giant star; f) a 
moderately high \SNR\ giant; g) a low \SNR\ giant; h) an emission line cool dwarf; i) a very 
cool star with molecular bands (using calibrated stellar parameters obtained via the \texttt{MADERA} pipeline (see DR6-2, Section 3.1). Error spectra are given as relative errors. A value of 
0.01 implies that a 1\%\ error in flux with an approximately Gaussian distribution is 
expected at this wavelength bin.}

\end{center}
\end{figure*}

The wavelength range of the \rave\  spectra is dominated by strong spectral lines: for the 
majority of stars, the dominant absorption features are due to the infrared calcium 
triplet, which in hot stars gives way to the Paschen series of hydrogen. There are also 
weaker metallic lines present for solar-type stars and molecular bands for the coolest 
stars. Within an absorption trough the flux is small, so shot noise is more significant 
in the middle of a line than in the adjacent continuum. Error levels also increase at 
wavelengths of airglow sky emission lines, which have to be subtracted during 
reduction. As a consequence, a single number, usually reported as \SNR, is not an 
adequate quantification of the observational errors associated with a given spectrum. 
For this reason, we provide error spectra that comprise uncertainties (“errors”) for 
each pixel of the spectrum. These are provided both for spectra prior to sky 
subtraction and for the final sky-subtracted ones (for details, see DR5, Section 4). 

The main contribution to the error spectrum is shot noise, which can be parametrized as 
$\SNR\ = g N_s / \sqrt{g N_u}$, where $N_u$ is the number of counts per pixel before sky 
subtraction, $N_s$ is its counterpart after the subtraction and the effective gain $g = 
0.416$~e$^-/$ADU (see DR5 for details). As explained above, the main source of the 
difference between $N_u$ and $N_s$ are airglow emission lines. So the relative flux 
errors increase within deep stellar absorption lines, such as the Ca~II infra-red 
triplet, and at positions of airglow lines. Note that subtraction of the latter in the 
sky subtracted spectra may be sub-optimal due to a rapid variability of sky airglow. 
Other contributions to the error spectrum are scattered light and imperfect
flat-field fringing removal, which typically contribute at a 0.8\%\ level, added in
quadrature. Finally, the resulting error spectra are smoothed with a window of a width equal to 3 pixels in the dispersion direction, which takes into account the noise correlation
between adjacent pixels.  

Error spectra are given as relative errors. A value of 0.01 implies that a 1\%\ error in 
flux with an expected approximately Gaussian distribution at this wavelength bin.  

\subsection{Signal to noise}\label{subsec:SNR}

As described in the previous paragraph, the errors on the normalized fluxes of the reduced  spectra {vary from pixel-to-pixel and it is problematic to represent a whole spectrum with a single value}.  On the other hand, {\it a priori\/} estimates of 
the \SNR\  per pixel are needed for the radial-velocity and stellar-parameter pipelines 
(see, \eg, DR2, section 3.4, DR3 section 2.2 or DR4, section 3.2). The corresponding 
values are reported in the respective catalog files (Section \ref{sec:FDR} and DR6-2, Section 7). 

\begin{figure}
\begin{center}
\plotone{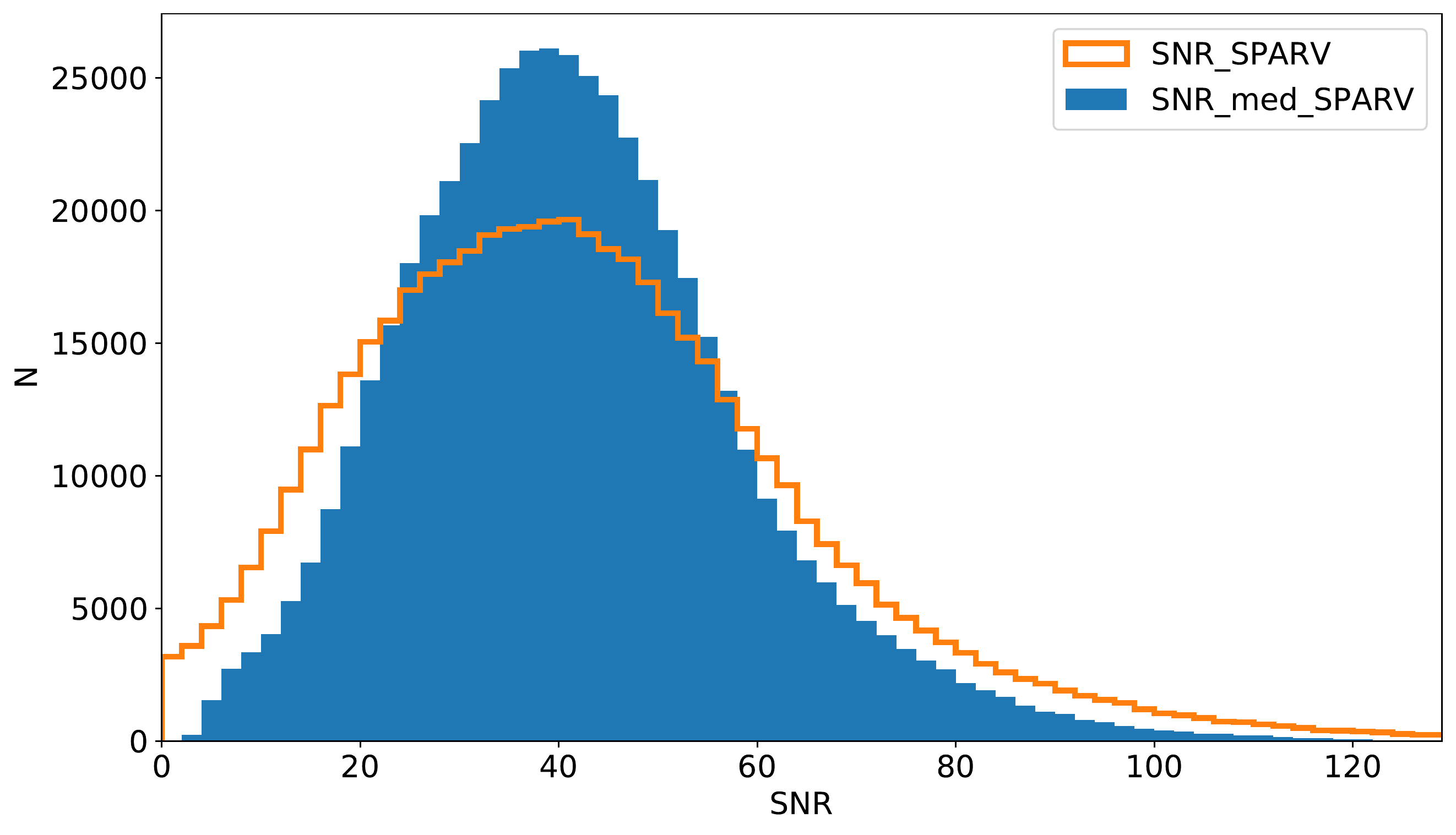}

\caption{\label{fig:SNR}
Distribution of the a posteriori signal-to-noise estimate \texttt{snr\_med\_sparv} and the a priori estimates \texttt{snr\_sparv} for all stellar spectra in the \rave\ DR6 database.}

\end{center}
\end{figure}

A better {\it a posteriori\/} estimate describing the quality of a spectrum, \eg, for data 
selection for a particular science application, is given by \texttt{snr\_med\_sparv}, 
defined as the inverse of the median of the error spectrum. \texttt{snr\_med\_sparv} 
scales on average with \texttt{snr\_sparv} in a {somewhat stronger than}  proportional manner (see Figure \ref{fig:SNR}). 
The median \texttt{snr\_med\_sparv} over the whole \texttt{SPARV} sample is $\approx 40$ (see Figure \ref{fig:SNR}).

\subsection{Morphology of \rave\ spectra}\label{subsec:samplespec}

A sample of typical  \rave\ spectra and their associated error spectra for a range of  targets is shown in 
Figure \ref{fig:sample_spec}, which, from top to bottom, represents: a) a high \SNR\ cool 
dwarf; b) a high \SNR\ warm dwarf; c) a high \SNR\ hot dwarf; d) a high \SNR\ red clump 
star; e) a high \SNR\ giant star; f) a moderately high \SNR\ giant; g) a low \SNR\ giant; 
h) an emission line cool dwarf; i) a very cool star with molecular bands. In all 
spectra other than those for the hot dwarf and the cool star with molecular bands, the Ca 
triplet can easily be recognized as the dominant feature. For hot dwarfs, the Ca triplet 
feature is replaced by strong and broad Paschen lines. Consequently, radial velocities 
can only be determined poorly for this class of stars, and atmospheric parameters (if 
at all convergent) are highly unreliable. The Ca triplet wavelength region also shows a 
considerable number of weak metal lines, which are used in Section 4 of paper DR6-2 to 
derive abundances of individual elements and \alphafe-ratios. These absorption lines are clearly visible in 
the high signal to noise cases and also in the moderate \SNR\ spectrum 6), but become 
difficult to discriminate against noise for \SNR\ of 20 and lower, as we demonstrate more 
quantitatively  in Section 6 of paper DR6-2.

\section{Spectral Classification}\label{subsec:Class}

The classification of \rave\  spectra was introduced in \cite{matijevic2012}. For this 
data release we modified the original classification scheme in order to simplify its 
use. Previously, the classification of \rave\  spectra was given as a series of 20 
flags for each spectrum. These flags represented the 20 nearest neighbors in the 
Locally Linear Embedding projected space and were ordered according to (decreasing)  
relative weights. In the revised version we first re-normalize all 20 
weights so they sum to unity  and then add all weights belonging to each flag. For 
example, in the case of a spectrum that has 13 normal star flags, 6 chromospherically 
active star flags and 1 binary star flag, we add 13 re-normalized weights for the 
normal stars and so on for the rest. This results in only three flags (for the flags 
and their occurrence see Table \ref{tab:Class}) plus their 
respective weights and enables the user to choose quantitatively  among the 
morphological types of spectra. It should be noted that the summed weights are not equal 
to the probability  that a spectrum belongs to a certain class but can be used as a proxy. In many cases all 20 original flags are of the same class so we only 
report a single flag with a single summed weight of 1.0. In cases where there were more 
than three different classes assigned to a single spectrum we report the first three, 
with the highest summed weights in decreasing order (the first one always being the 
largest). Consequently, the sum of the three weights $w1$, $w2$, and $w3$ is less than or 
equal to one. Of the 518,392 spectra in this release 490,959 (94.7\%) have the first of 
the flags with the value `n', i.e.\ they are classified as likely to be  normal stars. 

\begin{deluxetable}{c|lr|l}
\tablecaption{Description of the classification flags as described in the respective section of \cite{matijevic2012}.\label{tab:Class} }
\tablehead{\colhead{Label}  & \colhead{Description} & \colhead{$N_\mathrm{flag1}$} & \colhead{comment}}  
\startdata
        n & normal stars & $490,955$ & Section 4.1 \\
        o & hot stars ($\teff > 7000\,$K)& $5410$ & Section 4.1\\
        b & binary stars & $3123$ & Section 4.2\\
        d & cool dwarfs & $181$ & Section 4.3\\
        e & chromospheric emission lines /active stars & $6345$ & Section 4.3\\
        t & TiO band stars / cool giants & $5297$ & Section 4.4\\
        g & cool giants& $69$ & Section 4.5\\
        h & hot giants & $51$ & Section 4.5\\
        a & carbon stars & $271$ & Section 4.6\\
        p & peculiar stars & $82$ & Section 4.7\\
        c, w & problematic spectra & $6603$ & Section 4.8
\enddata
\end{deluxetable}

\begin{figure*}
\begin{center}
\plotone{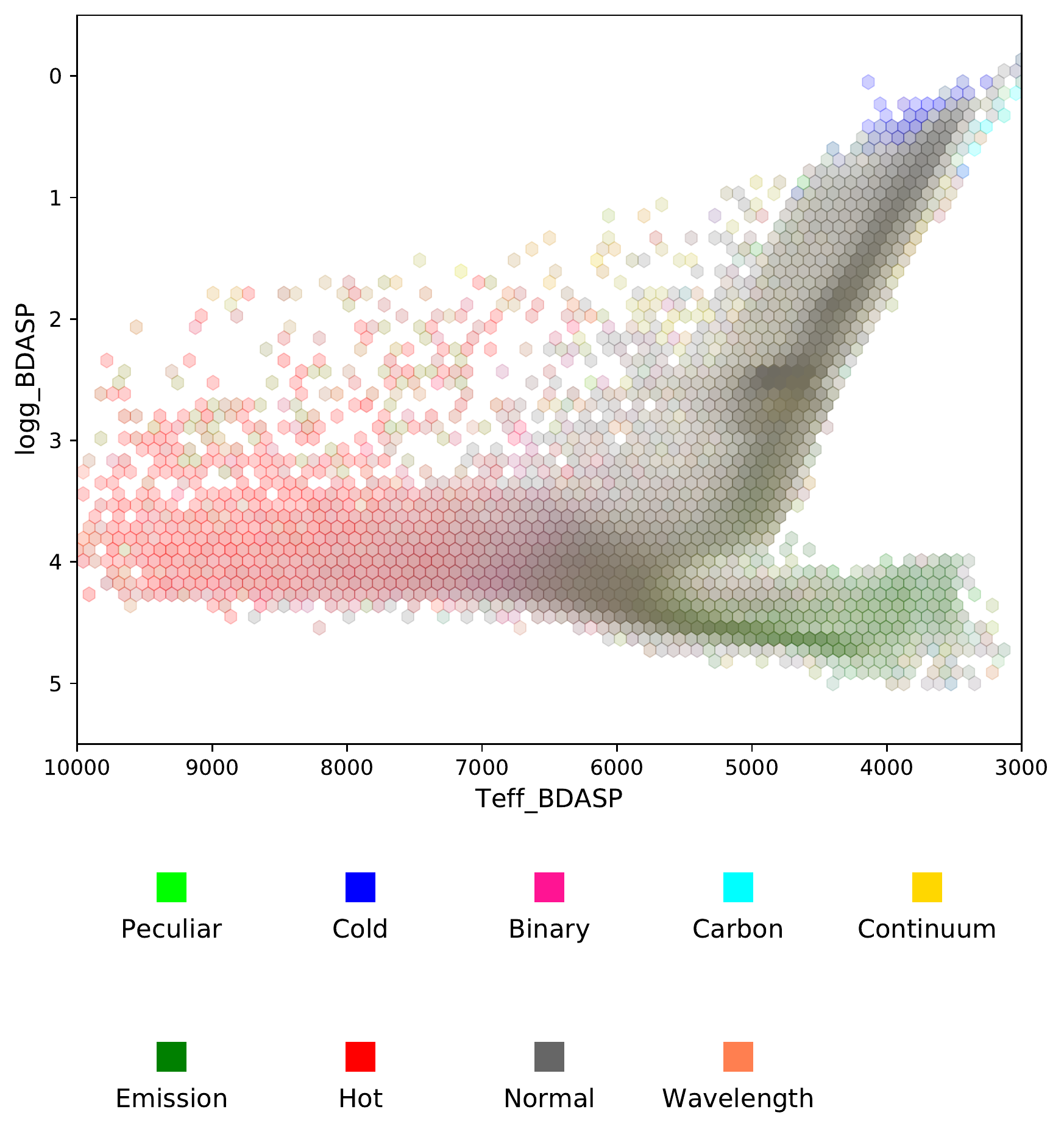}
\caption{\label{fig:kiel_class}
Kiel diagrams assuming :\teff\ and \logg\ from the \texttt{BDASP} pipeline (see DR6-2), color encoded by the average automated classification of the stars in the respective hexagon.}
\end{center}
\end{figure*}

We can both illustrate and verify the automated classification scheme by showing where stars of different classification lie in the 
\logg\ vs \teff\ plane (Kiel diagram, Figure \ref{fig:kiel_class}): The classification scheme nicely 
shows the transition to hot stars above a temperature of $\teff\approx 7000\,$K owing 
to the presence of strong Paschen line features, which dominate over the Ca triplet 
feature. On the main sequence, at effective temperatures below $5000\,$K, chromospheric emission lines become 
more prevalent in these cool and active stars \citep{zerjal2013}. 
At temperatures below $4000\,$K, molecular lines lead to a classification of the star 
as cool or as having carbon features, in particular near the tip of the giant 
branch. A slightly pinkish color in the sequence parallel to the main sequence for 
temperatures above 4500\,K also indicates a binary origin of stars in this part of 
the \logg\ vs \teff\ plane, for temperatures below 4500\,K, the emission line 
characteristics dominates the classification also in this part of the parallel 
sequence.

\section{Radial Velocities\label{sec:SPARV}}

\begin{figure}
\begin{center}
\plotone{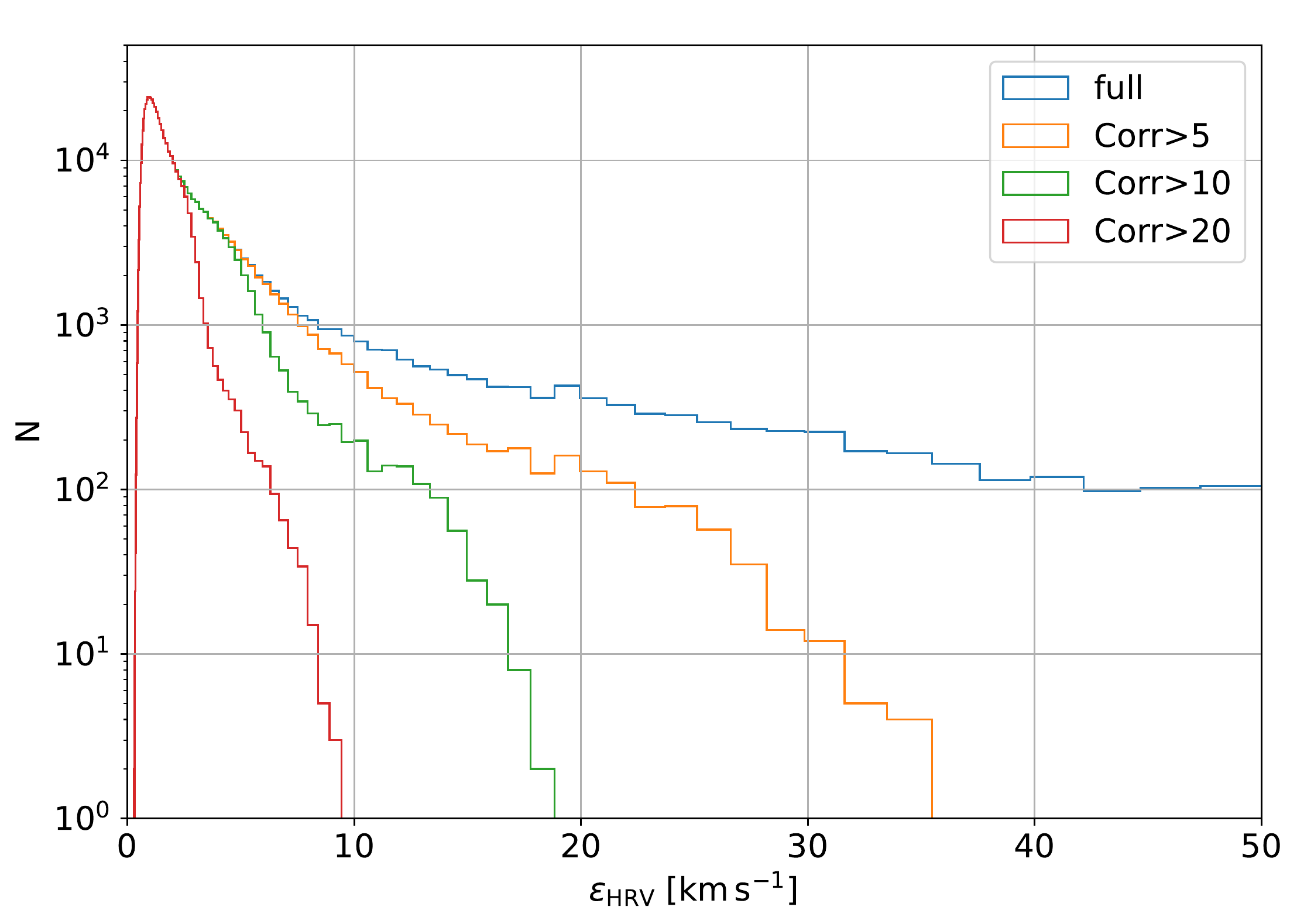}

\caption{\label{fig:r_error}
Distribution of the internal error estimate $\varepsilon_{\rm HRV}$ for the full sample and for subsamples with a correlation coefficient $R$ larger than 5, 10, and 20, respectively. Increasing the requirements on R strongly suppresses the tail of high-velocity errors.}
\end{center}
\end{figure}

Radial velocities (RVs) are derived with the pipeline \texttt{SPARV}  in a manner identical 
to that presented in DR4 and DR5, and as 
detailed in Section 2 of DR3. The spectra are cleaned in the spectral regions that are strongly affected 
by fringing (DR3 Section 2.4) and then matched to a grid of spectra discretized in 
\teff, \logg, \Mh, \alphafe, and stellar rotational velocity $v_\mathrm{rot}$, 
assuming a fixed microturbulence $\xi=2\,\kms$ . The underlying algorithm is a standard 
cross-correlation algorithm in Fourier space. The grid employs the  synthetic spectral 
library of \cite{munari2005} based on ATLAS 9 model atmospheres, and was extended with a 
finer grained spacing toward the densest region of the observed parameter space. The 
grid has $\Mh = -2.5, -2.0, -1.5, -1.0, -0.8, -0.6, -0.4, -0.2, 0.0, 0.2, 0.4, 
\mbox{and}\  0.5\,\dex$. {For stars cooler than $4500$~K the grid includes also molecular lines, while any influence of dust or chromospheric activity is neglected. The latter can be important in young cool stars descending toward the main sequence \citep{zerjal2013,zerjal2017}, so an increased template mismatch reflects in an increased RV error for such objects.} 
The process to match templates and thus derive RVs follows a 
two-step procedure: In a first step a provisional estimate of the radial velocity is 
obtained using a subset of only 10 template spectra. This first estimate typically 
results in RVs with an accuracy better than $5\,\kms$ and is used to put the spectrum in 
the zero velocity frame. Then a new template is created using a penalized chi-square 
method as described in DR2, which in turn is used to derive the final, more precise RV. 
To determine the zero point, the processing pipeline uses the available sky lines in the 
\rave\ wavelength window and fits a combination of a third-order polynomial and a constant function 
to the relation between sky radial velocity and fiber number (see DR3, Section 2.5). 
This fitting function defines the mean trend of zero-point offsets and provides the 
zero-point correction as a function of fiber number. The internal error is defined as 
the error in the determination of the maximum of the correlation function using IRAF 
\texttt{xcsao}. This procedure results in RVs with an internal error distribution 
peaking near $1\,\kms$ with a long tail towards higher RV errors probably owing to 
problematic spectra and/or variability from stellar binaries (see Figure \ref{fig:r_error}). 68\% of the sample has 
an internal accuracy better than $1.4\,\kms$ (see DR5). 

The RVs and their respective errors were confirmed by external observations and also by 
those targets that have repeat observations (see DR4 and DR5). The long tail in the 
repeat observations can be reduced by 90\% by applying quality criteria indicative of 
derived radial velocities with high confidence, namely 
$|\mathtt{correctionRV}|<10\,\kms$, $\sigma(\mathrm{RV})<8\,\kms$, and 
$\mathtt{correlationCoeff}>10$ \citep{kordopatis2013}.

RVs provided by \rave\  do not include corrections for gravitational redshift effects,  nor do they take into 
account any convective motions in the stellar atmosphere.  As 
discussed in DR2, our choice to omit these two contributions follows Resolution C1 of the IAU General Assembly in 
Manchester  \citep{rickman2001} and is consistent with the derived RVs reported by most other 
spectroscopic surveys, including Gaia DR2. The reader should however note that such an RV does not correspond 
to the line-of-sight component of the velocity of the stellar center of mass, which corresponds to the RV reported by the GALAH survey \citep{zwitter2018} and which is expected to be 
followed also by Gaia DR3. Typical values of gravitational redshifts are $+0.5 \pm 
0.2$~$\kms$ for dwarfs and $+0.1 \pm 0.1$~$\kms$ for giants, while convective shifts in 
the optical range are $-0.45 \pm 0.15$~$\kms$ for dwarfs and  $-0.3 \pm 0.2$~$\kms$ for 
giants \citep{zwitter2018}. As these values do not cancel exactly one should take care 
when studying the detailed internal dynamics of loosely bound stellar associations or 
streams where the reported  \rave\  RVs may exhibit systematic effects with spectral type at a level 
of $\sim 0.1\,\kms$.

\section{Validation of \rave\  DR6 parameters}\label{sec:Validation}

The data product of large surveys like \rave\ is always a compromise between the 
quality of the individual data entry and the area and depth of the survey. This applies 
to design decisions (like the applied exposure time/targeted \SNR) as well as to the 
decision which data to keep in the sample and which ones to exclude. Our policy for \rave\  
is to provide the maximum reasonable data volume possible, which allows the user to 
consider the tails of the distribution function. \textbf{The exact choice of the (sub)sample 
used for a particular science case has to be made by the user based on the 
criteria needed for the respective science application!} Here, we only can give some 
first guidelines/recommendations regarding the data downselection. For a description 
of the various parameters in the following paragraph, we refer to the tables in Section 
\ref{sec:FDR}.

Stars with $\texttt{correlation\_coeff} > 10$ 
have an internal velocity error distribution that peaks {near $\varepsilon_{\rm HRV}\approx 1-2$ \kms} with the tail toward very large velocity errors strongly suppressed compared to the uncut sample (see Figure \ref{fig:r_error}). For repeat measurements, such a sample features a small scatter in the repeat measurements of 
their heliocentric radial velocity. The distribution peaks near 
$0.0\,$\kms, and the tail toward very large velocity differences is reduced by 90\%, again 
compared to the uncut sample, indicative of a high confidence measurement (see below). We refer to 
the data set defined by these criteria as the core sample, or RV00.

{The reported internal RV errors reflect both, statistical uncertainty and systematics owing to a mismatch between observed and synthetic spectra. The \rave\ radial velocities and their uncertainties can be verified against  independent external sources (Section 6.1), providing a measure for the accuracy, or internally by considering repeat observations providing an estimate of the
precision (Section 6.2).   
}

We alert the reader that the above mentioned quality criteria were drawn under the assumption that \rave\ is used as a statistical sample. Should the \rave\ catalogs be used to identify individual candidates for followup studies (e.g. candidates for high radial velocities), additional criteria constraining the uncertainty of the measurement $\sigma(RV)$ and of the zero-point correction  ($\vert\texttt{correction\_rv}\vert$) should be applied. 

\begin{deluxetable*}{l|lrr}
\tablecaption{\rave\ subsamples used in this publication for validation and first science applications. \label{tab:ver_samples}}
\tablehead{\colhead{Sample} & \colhead{selection criteria} & \colhead{sample size} & \colhead{unique objects} }
\startdata
RV00 &  $\texttt{correlation\_coeff} > 10$ & 497,828 & 436,340\\
RV20 & RV \& \texttt{snr\_med\_sparv} $>20$ & 468,238 & 411,761\\
RV40 & RV \& \texttt{snr\_med\_sparv} $>40$ & 259,316 & 230,126\\
RV60 & RV \& \texttt{snr\_med\_sparv} $>60$ & 65,410 & 57,783\\
RV80 & RV \& \texttt{snr\_med\_sparv} $>80$ & 13,528 & 11,922
\enddata
\end{deluxetable*}

\subsection{Validation of Radial Velocities against Gaia DR2}\label{subsec:ext_val}

\begin{figure*}
\begin{center}
\plotone{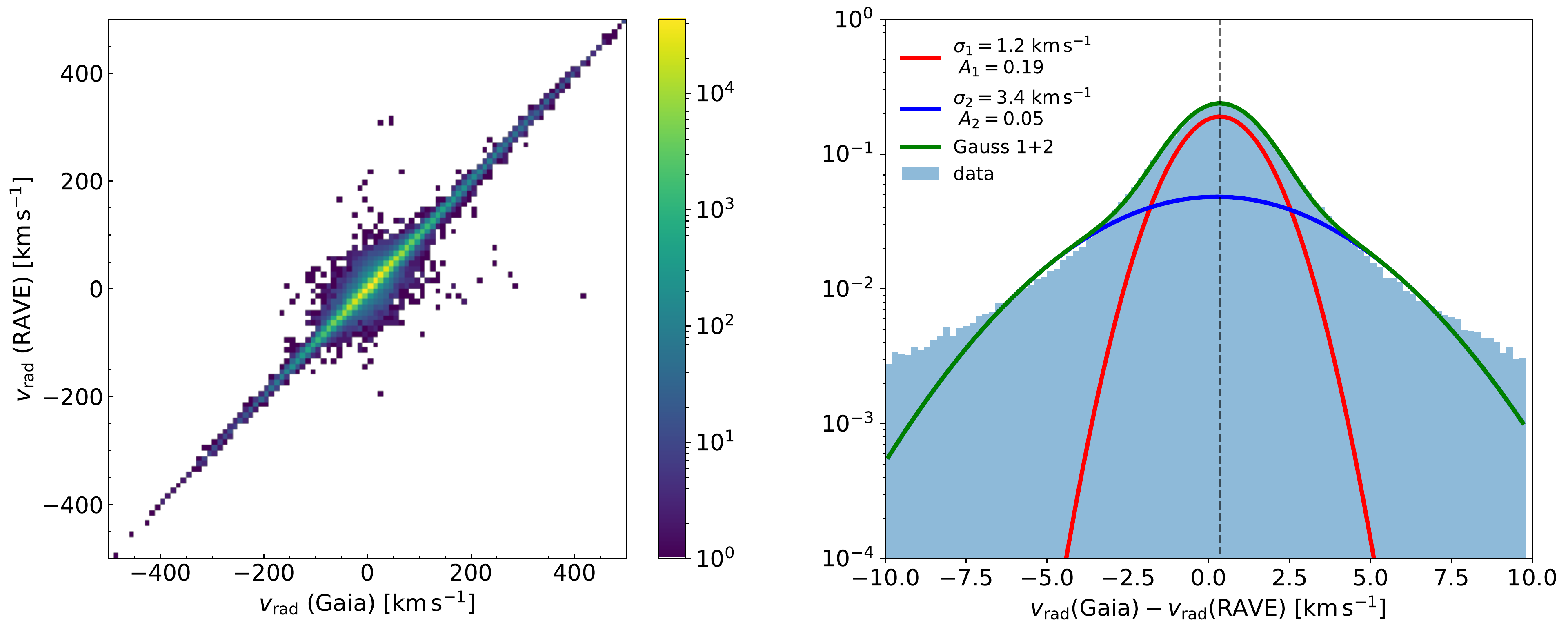}
\caption{\label{fig:RV_comp}
Left: Radial Velocity derived from \rave\  DR6 vs those from Gaia DR2. Right: 
Distribution of radial velocity differences between Gaia DR2 and \rave\  DR6. The green 
line compares this distribution function with a fit using two Gaussians with a standard 
deviation of $1.2\,\kms$ (red) and $3.4\,\kms$ (blue), respectively.  }
\end{center}
\end{figure*}

The accuracy of \rave\ radial velocities as compared to external observations was 
extensively discussed in DR3 Section 3.1 and DR4 Section 8.1. On 25 April 2018, the 2nd 
data release of the ESA mission Gaia was published \citep{GaiaDR2}, featuring radial 
velocities for some 7 million targets. The Radial Velocity Spectrometer (RVS) 
of Gaia also operates in the Ca triplet region, though at a somewhat higher resolution 
of $R=11,000$. The radial velocities of \rave\ DR5 and Gaia DR2 are compared in 
\cite{steinmetz2018}, showing a very good agreement between both data sets and also 
identifying a very small subset of \rave\ stars in DR5 with problematic wavelength 
calibration (almost exclusively stars at the edge of the field plate at observing periods 
with a high
rate of disabled fibers). These stars have been removed in DR6 (see Appendix \ref{sec:book}).
\rave\   DR6 and Gaia DR2 have 450,646 stars in common. This 
provides an opportunity to comprehensively compare \rave\ and Gaia radial velocities. 
Since \rave, however, provided the largest subset of targets for validating the Gaia 
pipeline \citep{sartoretti2018},  the two data sets are not fully independent. {Furthermore, the RVS of Gaia, covers the same spectral range at a similar resolving power, so that any spectral mismatches approximately cancel out each other.}  

Figure \ref{fig:RV_comp} compares the radial velocities published in Gaia DR2 with 
those presented here (RV00 sample). Overall this comparison confirms the excellent agreement between 
those two data sets. The velocity differences can well be matched with two Gaussians 
with standard deviations of { $1.2\,\kms$ and $3.4\,\kms$, respectively, plus an additional exponential tail towards higher velocity errors. We will discuss the possible origins of this behaviour in the next section.}

There is a
systematic offset of about $-0.32\,\kms$. The offset is also comparable to the offset 
found between Gaia DR2 and other ground-based spectroscopic surveys in a similar 
magnitude range, such as APOGEE \citep{sartoretti2018}, indicative that the source for 
this offset may at least partially be related to the radial velocity zero point of Gaia 
DR2.  The difference is also within the internal error estimates described above, 
errors as compared to external samples, and errors derived from a subset of stars with repeat 
observations. 

\begin{figure*}
\begin{center}
\plotone{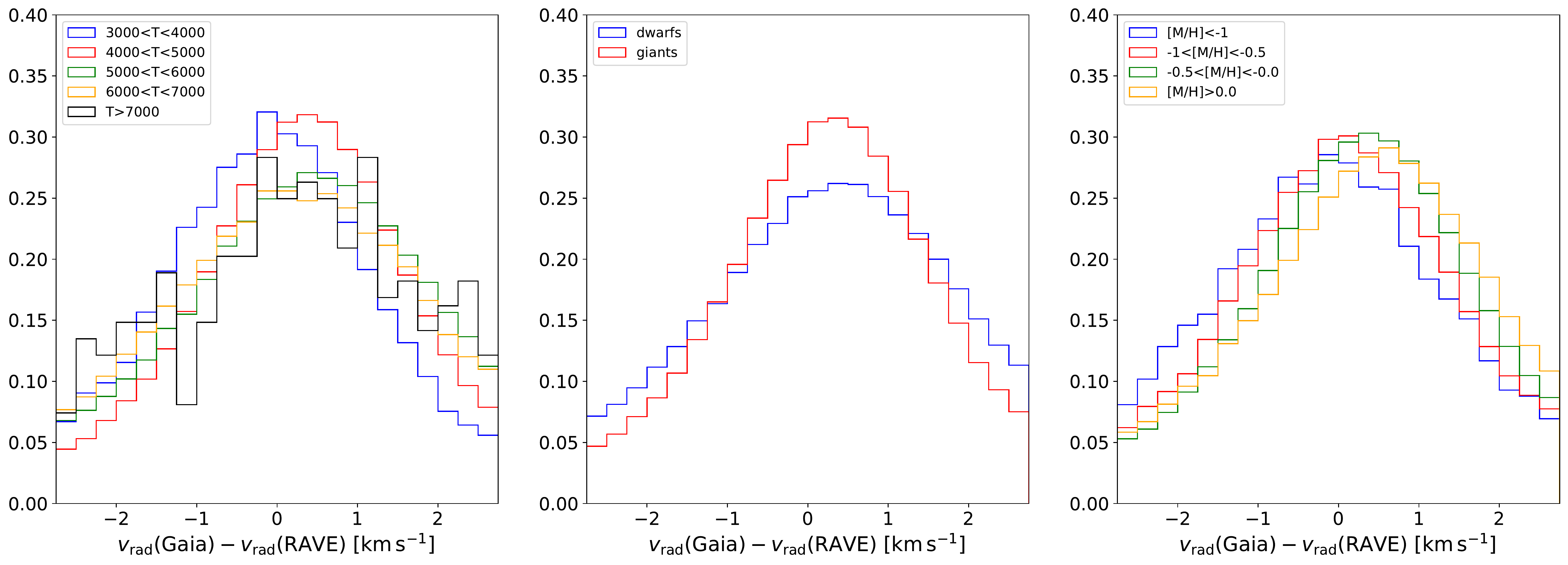}
\caption{\label{fig:RV_comp_parm}
Radial velocity differences between \rave\ DR6 and Gaia DR2 as a function of \texttt{BDASP} stellar parameters 
and \texttt{MADERA} \Mh. Left: temperature; Middle: giants ($\logg <3.5$) vs dwarfs ($\logg \geq 3.5$); Right: metallicity.}
\end{center}
\end{figure*}

A further analysis exhibits no systematic tendency of the RVs with \rave\   derived effective 
temperatures for stars with 4000\,K$<T_{\rm eff}<$7000\,K (Figure 
\ref{fig:RV_comp_parm}, left panel). Stars cooler than 4000\,K exhibit a somewhat 
smaller shift of $-0.1\,\kms$. For stars hotter than 7000\,K (a small subset of the 
\rave\   sample), the accuracy of the radial velocity deteriorates resulting in a 
larger systematic shift and a considerably increased spread, owing to the increasing 
dominance of broad Paschen lines at the expense of a less prominent Calcium triplet. 
With increasing \SNR, the prominence of the $1.2\,\kms$ Gaussian increases, while 
that of the $3.4\,\kms$ Gaussian decreases. A lower fraction of dwarf stars 
(log $g >3.5$) lies within the $1.1\,\kms$ Gaussian than for giant stars (Figure 
\ref{fig:RV_comp_parm}, middle panel).

There is a very mild tendency for the velocity shift between \rave\  and Gaia DR2 
to change with metallicity (Figure~\ref{fig:RV_comp_parm}, right panel). 
This effect amounts to about $0.5\,\kms$ between ${\rm \feh<-1}$ and  ${\rm \feh>0}$.

\subsection{Validation with repeat observations}\label{subsec:repeat_val}

A further way to validate the quality of the \rave\ data products is to compare the  
parameters derived for multiple observations of the same object (see Section 
\ref{subsec:Repeats}). In the following analysis we calculate for each 
star $k$ that has $N^k_\mathrm{repeat}>1$ observations{, that fulfills the quality 
threshold for the RV00 sample, and that has a match in the Gaia DR2 catalog. This corresponds to a total of $95 068$ spectra, or about 18\% of the total \rave\ database. We determine the uncertainty $\Delta RV$ in the radial velocity by three methods: 
\begin{itemize}
    \item From internal errors: $\Delta RV^k$ for star $k$ is randomly sampled assuming a normal distribution with a width corresponding to the internal error estimate $\epsilon^k_{HRV}$.  
    \item From repeat observations: The difference $\Delta RV_i^k$ between the radial velocity of star $k$ determined from observation $i$ ($1\leq i\leq 
N^k_\mathrm{repeat}$), $RV_i^k$,  and the mean $\overline{RV}^k$ for the 
respective repeat sequence. 
\item From comparison with Gaia DR2: The difference between the radial velocity from Gaia DR2 and that of \rave\ DR6.
\end{itemize}
}

We then analyze the distribution function over all stars  and observations and approximate it by  two Gaussians using a least-squares fit analogously to Section \ref{subsec:ext_val}.

\begin{figure*}
\begin{center}
\plotone{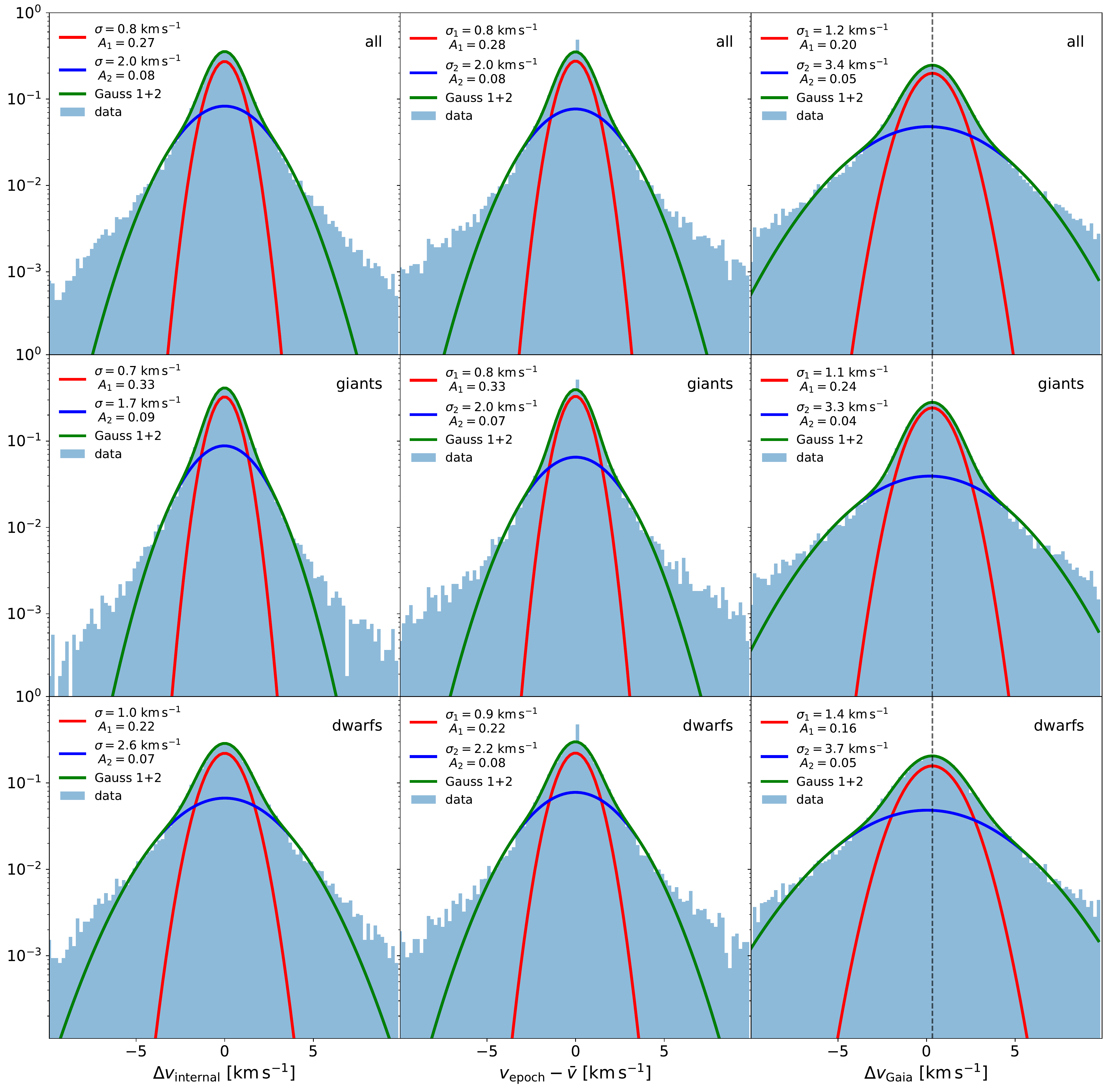}
\caption{\label{fig:repeat_RV} Difference 
in the radial velocity {for all stars (top row), dwarfs (medium row) and giants (bottom row) with 
more than one observation.  The left column shows the distribution of the expected velocity difference if the radial velocity $RV^k$ for each star $k$ is normally distributed with variance $\varepsilon^k_\mathrm{HRV}$. The middle column shows the difference of the radial velocity $RV^k_i$ of star $k$ measured at epoch $i$ against $\overline{RV}^k$, the radial velocity averaged over all epochs available for star $k$. The right column shows a comparison between the radial velocity measured by Gaia and the radial velocity measured by \rave. The red and blue curves correspond to the individual Gaussians of a two-Gaussian fit to the distribution, shown in green.
}}
\end{center}
\end{figure*}

{
The top row of Figure \ref{fig:repeat_RV} shows the 
distribution function in $RV$ for the aforementioned sample for three different methods in the left, middle, and right column, respectively. Furthermore we show the same analysis separated into giants (52405 spectra, middle row) and dwarfs (42663 objects, bottom row)}. In each panel a fit of the distribution with two Gaussians is shown.

{The comparison between the repeat sequence and the internal error distributions gives very consistent results, basically resulting in very similar values for both fitting Gaussian for each of the three samples (all, giants, dwarfs).

The comparison with Gaia RVs finds identical fitting parameters as in Figure \ref{fig:RV_comp} indicating no systematic difference between the repeat sample studied here and the full RV00 data set. The width of the narrower Gaussians is a factor of $1.4$ larger, the factor of the wider Gaussian is typically somewhat higher. This behavior is in very good agreement with our expectation: for two independent observations of similar uncertainty (which is what we would expect for Gaia and \rave\ considering the comparable resolution and \SNR), the errors should add in quadrature resulting in a $\sqrt{2}$ wider distribution function\footnote{note that for the repeat sequence the difference at on epoch to the mean is taken rather than the difference between the velocity at two observation epoch. In the latter case we get the same $\sqrt{2}$ factor}.
}

{The fit by two Gaussians should be merely seen as a simple model to approximate a distribution function which cannot be approximated by a single normal distribution but exhibits considerable non-normal wings. The reason for this wings are manifold and include: (i) poorer fits in general for dwarf than for giants; (ii) systematic decrease in accuracy towards higher temperatures owing to the less prominent Ca triplet feature; (iii) objects with intrinsically variable RVs, as the observation epochs of RAVE and Gaia DR2 data are between a few days and up to 12 years apart, (iv) varying presence of emission line features for active stars that can vary on similar time scales; (v) objects with a large mismatch between the observed spectrum and its best synthetic counterparts. The latter can be a consequence of inadequate modelling of certain types of spectra but also poor observing conditions or suboptimal instrument performance. We note that the latter is more common for stars observed through fiber numbers 1--2, or 145--150, i.e.\ the ones at the extreme edges of the CCD. Consistent with the assumptions (iv) and (v) we also find the wings of the distribution function to be less pronounced and the wider Gaussian to be somewhat suppressed for subsamples limited to high \SNR.

Based on this analysis we conclude that 68\%\ of the radial velocities on \rave\ DR6 have a velocity accuracy better than $1.4\,\kms$, while 95\%\ of the objects have radial velocities better than $4.0\,\kms$.

}

\begin{deluxetable*}{rlccll}
\tablecaption{\label{tab:DR6_SPARV}\texttt{DR6\_SPARV} catalog description} 
\tablehead{\colhead{Col}  &  \colhead{Format}   &   \colhead{Units} &   \colhead{NULL} & \colhead{Label}      &               \colhead{Explanations}}
\startdata
1       & char     & -          & N     & \texttt{rave\_obs\_id}             & \rave\ spectrum designation\tablenotemark{a}                                                    \\
2       & char     & -          & N     & \texttt{raveid  }                    & \rave\ target designation\tablenotemark{b}                                               \\
3       & char     & -          & N     & \texttt{objectid}                   & object identifier used in input catalog\tablenotemark{c}                                               \\
4       & float        & \kms       & N     & \texttt{hrv\_sparv  }                       & Heliocentric radial velocity (HRV)                                          \\
5       & float        & \kms       & N     & \texttt{hrv\_error\_sparv}                        & HRV error                                                              \\
6       & float        & -     & N     & \texttt{correlation\_coeff\_sparv}                &  Tonry-Davis $R$ correlation coefficient                                                           \\
7       & float        & -        & N     & \texttt{correction\_rv\_sparv}                &  zero-point correction of the HRV                                                            \\
8       & float        &  -       & N     & \texttt{chisq\_sparv}               &  $\chi^2$ of the \texttt{SPARV} pipeline                                                           \\
9 	  & float        &  -             & N     & \texttt{snr\_med\_sparv}			& median \SNR\ \tablenotemark{d}  \\
\enddata
\tablenotetext{a}{Observation date, field name, fibre number.}\tablenotetext{b}{J2000 GCS RA and Dec.}\tablenotetext{c}{Tycho-2, SSS, DENIS.} \tablenotetext{d}{as derived from \texttt{SPARV}, see Section \ref{subsec:SNR}.}
\end{deluxetable*}

\begin{deluxetable*}{rlccll}
\tablecaption{\label{tab:DR6_ObsData}\texttt{DR6\_ObsData} catalog description.} 
\tablehead{\colhead{Col}  &  \colhead{Format}   &   \colhead{Units} &   \colhead{NULL} & \colhead{Label}      &               \colhead{Explanations}}
\startdata
1       & char     & -          & N     & \texttt{rave\_obs\_id}   & \rave\ spectrum designation                                                    \\
2       & char     & deg        & N     & \texttt{ra\_input}       & RA in input catalog\\
3       & char     & deg        & N     & \texttt{dec\_input}      & Dec in input catalog\\
4       & char     & -          & N     & \texttt{field}           & field denotator, composite of obsdate and fieldname\\
5       & int      & -          & N     & \texttt{obsdate}         & Observation date yyyymmdd\\
6       & char     & -          & N     & \texttt{fieldname}       & name of the field: RA and Dec of field center\\
7       & char     & -          & N     & \texttt{fibernumber}       & Number of optical fiber [1,150]\\
8       & char     & - 	        & N     & \texttt{ut\_start} & exposure start in Coordinated Universal Time \\
9       & char     & - 	        & N     & \texttt{ut\_end} & exposure end in Coordinated Universal Time \\
10     & char       &   -      & N & \texttt{lst\_start} & exposure start in Local Sidereal Time \\
11     & char       &   -      & N & \texttt{lst\_end} & exposure end  in Local Sidereal Time \\
12       & int    & s          & N     & \texttt{exposure\_time}       & total exposure time\\
13      & char     & deg        & N     & \texttt{ra\_field}       & RA field center\\
14      & char     & deg        & N     & \texttt{dec\_field}      & Dec field center\\
15      & int          & -          & N     & \texttt{platenumber}                 & Number of field plate [1..3]                                           \\
16      & float          &  -         & N     & \texttt{airmass}                 &  Airmass \\
17      & float          &  -         & Y     & \texttt{lunar\_phase}                 & Lunar phase \\
18       & char     & -          & N     & \texttt{healpix4096}                      & HEALPix value\tablenotemark{a}     \\
19 	& int	    	& -	    & Y       & \texttt{cluster\_flag}		     & 1: targeted observation, NULL: otherwise \\
21 	& int	    	& -	    & Y       & \texttt{footprint\_flag}		     & 1: star in the \rave\ footprint, NULL: otherwise
\enddata
\tablenotetext{a}{Hierarchical Equal-Area iso-Latitude Pixelisation (HEALPix) values were 
computed using the resolution parameter $N_{\rm side} = 4096$ (resolution index of 12) and 
the NESTED numbering scheme. Any lower-resolution index HEALPix value
can be computed from the given one by dividing it by $4^{(12 - n)}$, where $n < 12$
is the desired resolution index.} 
\end{deluxetable*}
\begin{deluxetable*}{rlccll}
\tablecaption{\label{tab:DR6_Class}\texttt{DR6\_Class} catalog description.} 
\tablehead{\colhead{Col}  &  \colhead{Format}   &   \colhead{Units} &   \colhead{NULL} & \colhead{Label}      &               \colhead{Explanations}}

\startdata
1       & char     & -          & N     & \texttt{rave\_obs\_id}                 & \rave\ spectrum designation                                                     \\
2       & char        & -       & N     & \texttt{flag1\_class}                        & Primary flag                                          \\
3       & char        & -       & Y     & \texttt{flag2\_class}                         & Secondary flag                                           \\
4       & char        & -       & Y     & \texttt{flag3\_class}                         & Tertiary flag                                           \\
5       & float        & -       & N     & \texttt{w1\_class}                        & Weight associated with primary flag                                                              \\
6       & float        & -       & Y     & \texttt{w2\_class}                        & Weight associated with secondary flag                                                              \\
7       & float        & -       & Y     & \texttt{w3\_class}                        & Weight associated with tertiary flag                                    
\enddata
\end{deluxetable*}
\begin{deluxetable*}{rlcll}
\tablecaption{\label{tab:DR6_Repeats}\texttt{DR6\_Repeats} catalog description.} 
\tablehead{\colhead{Col}  &  \colhead{Format}   &    \colhead{NULL} & \colhead{Label}      &               \colhead{Explanations}}
\startdata
1       & char           & N     & \texttt{raveid}                 & Unique object identifier                                                     \\
2       & int           & N     & \texttt{n\_repeats}               & number of repeat observations (between 1 and 13)                                          \\
3-       & char             & Y     & \texttt{rave\_obs\_id1} - & unique spectrum identifiers\\ 
15&&& \texttt{rave\_obs\_id13}                         & for all repeat observations                \enddata
\end{deluxetable*}
\section{The sixth \rave\ public data release: catalog presentation I}\label{sec:FDR}

\rave\ DR6 spectra and derived quantities are made available through a data base 
accessible via \texttt{doi:10.17876/rave/dr.6/}. Since key words and unquoted identifiers are case insensitive, in SQL, in general lower case identifiers are used in the data base. The two main identifiers are 
\texttt{rave\_obs\_id} and \texttt{raveid}: the former, \texttt{rave\_obs\_id}, is the 
unique identifier denoting the observation of a particular spectrum -- the name is a composite of the observing date, field name, and fiber number allocated to the star on that occasion. 

\texttt{raveid} is the unique identifier of the target star, the name being a 
composite of the targets Galactic coordinates in the J2000.0 system. Consequently, 
objects that have several observations have the same \texttt{raveid} for all, but differ 
in their \texttt{rave\_obs\_id}. The data base contains also a considerable number of 
auxiliary parameters that can be employed to further scrutinize the specifics of the  
reductions using the various pipelines. These variables are described on the 
aforementioned website. Furthermore, ample information regarding cross-identification 
with other catalogs is given.

For convenience we also provide a set of FITS, CSV, and HDF files of
the overall \rave\ catalog, featuring key variables sufficient for the 
majority of applications of the \rave\ survey. These data are organized in 16 
files according to the pipeline employed; the content for 6 of these files is briefly 
described in the following paragraphs and associated tables, for the remining 10 we refer to paper DR6-2. We avoid duplication 
of variable entries in the different files, with the exception of 
\texttt{rave\_obs\_id}, which can be used to link the contents of the various catalogs.

\subsection{The \rave\ DR6 catalog of spectra}

\rave\ spectra and error spectra are available via the data base on the \rave\ webpage\footnote{\texttt{http://www.rave-survey.org}} 
\mbox{(DR6\_Spectra, \texttt{doi:10.17876/rave/dr.6/019})}. 
Spectra are made available in FITS files with a name based on 
their \texttt{rave\_obs\_id} containing (i) the actual wavelength-calibrated and flux 
normalized spectrum and (ii) the associated error spectrum (for example, see  
Figure~\ref{fig:sample_spec}). 

Only spectra that successfully passed the 
\texttt{SPARV} pipeline (\ie, where a radial velocity can be derived) are added to the 
data base.

\subsection{The \rave\ DR6 catalog of radial velocities}

The \texttt{DR6\_SPARV} table (\texttt{doi:10.17876/rave/dr.6/001}) should be seen as the 
master file of the \rave\ DR6 data 
release. It contains all observations and objects for which a spectrum can be found in 
the spectral data base, contains all observations for which the pipeline converged to 
provide a radial velocity, and is the sample of spectra that served as input for further 
analysis pipelines, \eg, those to derive stellar atmospheric parameters or abundances. 

\texttt{DR6\_SPARV} contains the heliocentric radial velocity, information on the 
zero-point calibration using sky lines (see Section \ref{sec:SPARV}) and convergence 
information of the pipeline (Table \ref{tab:DR6_SPARV}). In the \rave\ data base, 
additional information such as stellar parameters from matching templates is provided 
(\texttt{DR6\_SPARV\_aux}, \texttt{doi:10.17876/rave/dr.6/002}). These data should, if 
at all, be used with care for further 
astrophysical applications, as they are subject to complicated biases (see DR4, Section 
4.4). 

\subsection{The \rave\  DR6 catalog of diagnostic data}

\texttt{DR6\_ObsData} (\texttt{doi:10.17876/rave/dr.6/003}) contains helpful 
diagnostic information regarding the \rave\ data 
and the derived data products (Table \ref{tab:DR6_ObsData}). This includes, \eg, the 
observing date, exposure time, fiber number, field plate used, number of arc lines used 
for the wave length calibration, the coordinates of the field plate center, phase of 
Moon, and healpix coordinates.

\subsection{The \rave\  DR6 catalog of classification}

Results of the automated classification (Section \ref{subsec:Class}) are assembled in the 
\texttt{DR6\_Class} file \\
(\texttt{doi:10.17876/rave/dr.6/004}, Table \ref{tab:DR6_Class}),
giving up to 3 classification flags and their relative weights.

\subsection{The \rave\  DR6 catalog of repeat observations}

To enable an easy analysis of stars with more than one observation date, the 
\texttt{DR6\_Repeats} file \\
(\texttt{doi:10.17876/rave/dr.6/005}, Table 
\ref{tab:DR6_Repeats}) features the 
\texttt{raveid} to  identify the target uniquely, the number of revisits 
\texttt{n\_Repeats}, and the respective \texttt{rave\_obs\_id}s of all observations of 
that particular target (for a detailed analysis, see Section~\ref{subsec:Repeats}).

\subsection{Cross match of \rave\  DR6 with Gaia DR2 and other catalogs}

The \rave\ DR6 data release is complemented by two files cross-matching \rave\ DR6 with Gaia DR2 
(\texttt{DR6\_GaiaDR2}, \texttt{doi:10.17876/rave/dr.6/015}) and with a suite of other catalogs 
including Tycho-2, 2MASS, WISE, APASS9, and SKYMAPPER (\texttt{DR6\_XMatch}, 
\texttt{doi:10.17876/rave/dr.6/016}).

\section{Summary and conclusions}\label{sec:conclude}
The \rave\ final data release concludes a more-than-15 year effort to provide a 
homogeneous data set for Galactic archaeology studies. \rave\ DR6 presents spectra and 
radial velocities for individual stars in the magnitude range $9<I<12$\,mag. The spectra 
cover a wavelength range of $8410-8795\,$\AA\ at an average resolution of $R\sim7500$. The 
\rave\ catalog can be accessed via \texttt{doi:10.17876/rave/dr.6/001}. The typical 
\SNR\ of a \rave\ star is 40, and the typical uncertainty in radial velocity is 
$< 2$\,\kms.  Catalogs containing observing 
statistics, repeat observations, an automated classification scheme, and cross matches with the Gaia DR2 and other catalogs 
such as 2MASS, DENIS, HIPPARCOS, TYCHO2, WISE, SKYMAPPER, and APASS9 complement the 
\rave\ final data release. Accompanying derived data products
such as stellar parameters, chemical abundances, and distances as well as some science applications are presented in Paper DR6-2.

\acknowledgments Major scientific projects like the \rave\ survey are made possible by the contributions of many, in particular those of graduate students and postdocs. This final data release is published in memory of one of the first and most active student participants in \rave\, Gregory R. Ruchti (1980 - 2019), whose life was taken far too early.  His enthusiasm and  dedication were key elements of the success  of the \rave\ collaboration and his  contributions live on in the discoveries that are enabled by the \rave\ data.

Funding for \rave\  has been provided by: the Leibniz-Institut f\"{u}r Astrophysik 
Potsdam (AIP); the Australian Astronomical Observatory;  the Australian National 
University; the Australian Research Council; the French National Research Agency (Programme National Cosmology et Galaxies (PNCG) of CNRS/INSU with INP and IN2P3, co-funded by CEA and CNES); the 
German Research Foundation (SPP 1177 and SFB 881: Project-ID 138713538); the European Research Council 
(ERC-StG 240271 Galactica); the Istituto Nazionale di Astrofisica at Padova; The Johns 
Hopkins University; the National Science Foundation of the USA (AST-0908326); the W. M. 
Keck foundation; the Macquarie University; the Netherlands Research School for Astronomy; 
the Natural Sciences and Engineering Research Council of Canada; the Slovenian Research 
Agency (research core funding no. P1-0188); the Swiss National Science Foundation; the Science \& Technology Facilities 
Council of the UK; Opticon; Strasbourg Observatory; and the Universities of Basel, 
Groningen, Heidelberg and Sydney. PJM is supported by grant 2017-03721 from the Swedish Research Council. LC is the recipient of the ARC Future Fellowship 
FT160100402. RAG acknowledges the support from the PLATO CNES grant. SM would like to acknowledge support from the Spanish Ministry with the Ramon y Cajal fellowship number RYC-2015-17697. MV acknowledges support of the Deutsche Forschungsgemeinschaft 
(DFG, project number: 428473034). MS thanks Research School of Astronomy \& Astrophysics in Canberra for 
support through a Distinguished Visitor Fellowship. RFGW thanks the Kavli Institute for 
Theoretical Physics and the Simons Foundation for support as a Simons Distinguished 
Visiting Scholar. This research was supported in part by the National Science Foundation 
under Grant No. NSF PHY-1748958 to KITP.

This work has made use of data from the European Space Agency (ESA) mission
{\it Gaia} (\url{https://www.cosmos.esa.int/gaia}), processed by the {\it Gaia}
Data Processing and Analysis Consortium (DPAC,
\url{https://www.cosmos.esa.int/web/gaia/dpac/consortium}). Funding for the DPAC
has been provided by national institutions, in particular the institutions
participating in the {\it Gaia} Multilateral Agreement.

\software{HEALPix \citep{healpix}, IRAF \citep{iraf}, Matplotlib \citep{matplotlib}, numpy \citep{numpy}, pandas \citep{pandas}, RVSAO \citep{rvsao}}

\appendix

\section{Bookkeeping of \rave\ observations}\label{sec:book}

In total there are 7041 \rave\ DR5 spectra that are not in this final data release. These fall into to the following two categories:

\begin{enumerate}
    \item Fewer than 4 arc lines were available for wavelength calibrations. This condition mainly occurs near the edges of the field plate owing to the fast focal ratio of the spectrograph camera when many fibers have been broken \citep{steinmetz2018} (687 spectra).
    \item Cases where the processing failed or the corresponding error spectrum could not be computed (6434 spectra).
\end{enumerate}

Furthermore, 4727 spectra were added that are not in \rave\ DR5, mainly corresponding to targeted observations in the context of the Aquarius substructure \citep{williams2011}.

\clearpage
\bibliographystyle{aasjournal}
\bibliography{sample}

\end{document}